# Reduced Recombination *via* Tunable Surface Fields in Perovskite Solar Cells

Dane W. deQuilettes,[1] Jason Jungwan Yoo,[2,3] Roberto Brenes,[4] Felix Utama Kosasih,[5] Madeleine Laitz,[4] Benjia Dak Dou,[1] Daniel J. Graham,[6] Kevin Ho,[7] Seong Sik Shin,[3*] Caterina Ducati,[5] Moungi Bawendi,[2*] Vladimir Bulović[1,4*]

[1] Research Laboratory of Electronics, Massachusetts Institute of Technology, 77 Massachusetts Avenue, Cambridge, Massachusetts, USA

[2] Department of Chemistry, Massachusetts Institute of Technology, 77 Massachusetts Avenue, Cambridge, Massachusetts, USA

[3] Division of Advanced Materials, Korea Research Institute of Chemical Technology, Daejeon, Republic of Korea.

[4] Department of Electrical Engineering and Computer Science, Massachusetts Institute of Technology, Cambridge, MA, USA.

[5] Department of Materials Science and Metallurgy, University of Cambridge, 27 Charles Babbage Road, Cambridge CB3 0FS, UK.

[6] Department of Bioengineering, University of Washington, Box 351653, Seattle, Washington, USA

[7] Department of Chemistry, University of Washington, Box 351700, Seattle, Washington, USA

**Corresponding Authors:** bulovic@mit.edu, mgb@mit.edu, and sss85@krict.re.kr

**Abstract**

The ability to reduce energy loss at semiconductor surfaces through passivation or surface field engineering has become an essential step in the manufacturing of efficient photovoltaic (PV) and optoelectronic devices.[1] Similarly, surface modification of emerging halide perovskites with quasi-2D heterostructures is now ubiquitous to achieve PV power conversion efficiencies (PCEs) > 22% and has enabled single-junction PV devices to reach 25.7%, yet a fundamental understanding to how these treatments function is still generally lacking.[2] This has established a bottleneck for maximizing beneficial improvements as no concrete selection and design rules currently exist.[2] Here we uncover a new type of tunable passivation strategy and mechanism found in perovskite PV devices that were the first to reach the > 25% PCE milestone,[3] which is enabled by surface treating a bulk perovskite layer with hexylammonium bromide (HABr). We uncover the simultaneous formation of an iodide-rich 2D layer along with a Br halide gradient achieved through partial halide exchange that extends from defective surfaces and grain boundaries[4] into the bulk layer. We demonstrate and directly visualize the tunability of both the 2D layer thickness, halide gradient, and band structure using a unique combination of depth-sensitive nanoscale characterization techniques. We show that the optimization of this interface can extend the charge carrier lifetime to values > 30 $\mu$s, which is the longest reported for a direct bandgap semiconductor (GaAs, InP, CdTe) over the past 50 years.[5] Furthermore, we show that this heterostructure is well suited for a host of optoelectronic devices where we achieve a new benchmark for perovskite/charge transport layer surface recombination velocity with values < 7 cm s$^{-1}$. Importantly, this work reveals an entirely new strategy and knob for optimizing and tuning recombination and charge transport at semiconductor interfaces and will likely establish new frontiers in achieving the next set of perovskite device performance records, addressing long-term operational stability issues, and bolstering the commercialization readiness of perovskite technology.[2]

**Ultralong Carrier Lifetimes with HABr Treatment**

Here, we study mixed-cation lead halide perovskite (FAPbI$_3$)$_{1-x}$(MAPbBr$_3$)$_x$ layers (see SI for preparation details) post-treated with varying concentrations (10-50 mM) of hexylammonium bromide (HABr) in chloroform (Figure 1a). Previously, this treatment has enabled two certified power conversion efficiency (PCE) records of 24.2 and 25.2%,[3,6] although its role in modifying the structure and function of PV devices has yet to be revealed. Therefore, uncovering the working mechanism offers the tantalizing prospect of gaining unprecedented control of 2D/3D interfaces to yield even higher performing optoelectronic devices. Indeed, by directly visualizing changes in nanoscale structure and composition as a function HABr concentration, we are able to demonstrate fine-tuning of local charge carrier distributions to achieve significant reductions in non-radiative recombination. To begin with, Figure 1b shows cross-sectional bright field transmission electron microscopy (TEM) images of thin 2D layers formed on top of the 3D bulk perovskite after treatment with 10 and 50 mM of HABr.[6] We note that there is a slight increase in the 2D layer thickness as the HABr concentration increases, highlighting that aspects of this heterostructure can already be tuned.

In order to assess the quality of these layers with varying HABr concentration, we determine the charge carrier lifetimes through photoluminescence (PL) decay measurements.[7,8] Minority carrier lifetime is a hallmark property of all semiconductors and is often referred to as the most important electronic property for solar cells.[9-11] Indeed, the success of commercial photovoltaics (PVs) as well as steady improvements in research-scale cells are often attributed to longer carrier lifetimes due to better surface passivation strategies.[1,12] Unpassivated bulk perovskite films deposited on glass now routinely demonstrate lifetimes $\sim 1$ $\mu$s,[8,13] which can be further improved to values as high as 8 $\mu$s with surface passivating molecules such as $n$-trioctylphosphine oxide (TOPO)[14] or up to 18 $\mu$s with solvent additives.[13] The carrier lifetimes of bare 2D/3D heterostructures on glass are typically only slightly longer than the as-grown 3D bulk layers with lifetimes ranging from hundreds of ns to a few $\mu$s.[15,16] In Figure 1c, we observe an increase in PL decay time as a function of HABr concentration, suggesting that the surface recombination kinetics are highly sensitive to concentration. To accurately determine the recombination rate values of these samples, we globally fit time-resolved PL (TRPL) decay measurements at a range of excitation fluences when excited from the glass side using the standard kinetic rate equation[7] (see SI for fitting details). For the control sample, we report an effective lifetime, $\tau_{eff}$, of 6.1 ± 0.9 $\mu$s (95% confidence interval (CI), $1/\tau_{eff} = k_1 = 1.6 ± 0.2$ x10$^5$ s$^{-1}$), consistent with a high-quality as-grown layer (see Figure S1).[17] For a champion 50 mM HABr treated sample (Figure 1d), we report an impressive lifetime of 32.7 ± 7.9 $\mu$s (95% CI, $k_1 = 3.1 ± 0.9$ x10$^4$ s$^{-1}$), with a decay time as long as 134 $\mu$s if we analyze the tail of the PL trace (Figure S2) as done in other reports.[18,19] This $\tau_{eff}$ value is higher than all previous reports for 2D/3D architectures, surpasses all previous perovskite records (see Figure S3), and is the longest value ever reported for high quality direct bandgap semiconductors (GaAs, InP, and CdTe) spanning the past 50 years (Figure 1f and see SI Tables 2-5) when measured using standard TRPL techniques. In fact, this value is close to the radiative lifetime limit ($\sim \geq 60$ $\mu$s) set by the background doping density (see SI for details). If PCE is correlated with minority carrier lifetime as it is in GaAs,[11] then perovskites may have the potential to surpass the 29.1% PCE single junction record set by GaAs. We note that the effective internal second-order recombination rate constant ($k_{2,eff}^{int}$ = 5.5x10$^{-11}$ cm$^3$ s$^{-1}$) for the 50 mM HABr treated sample is lower than the control sample ($k_2^{int}$ = 7.4x10$^{-10}$ cm$^3$ s$^{-1}$, see Figure S1 and S4) and other reported $k_2^{int}$ values,[7,20] suggesting slowed radiative recombination, which we explore in more detail below. In addition, consistent with a reduction in non-radiative loss, Figure S5 shows an improvement in the PL quantum efficiency (PLQE) of a typical treated sample compared to the control.

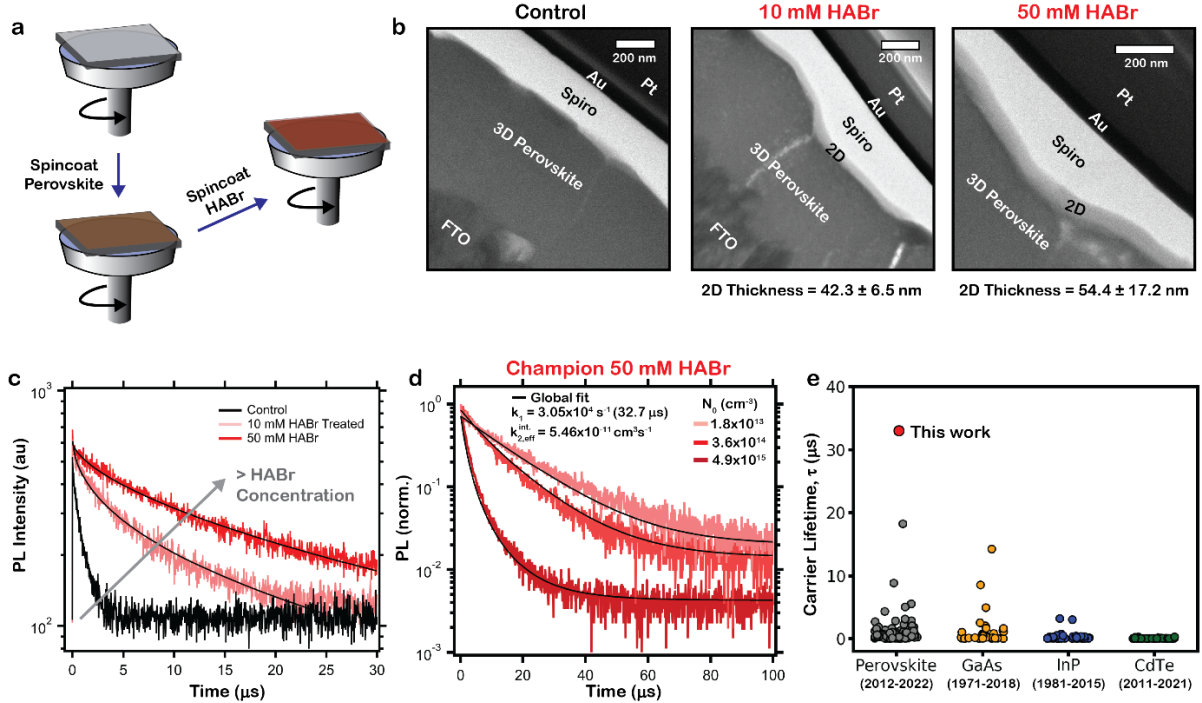

**Figure 1. a,** Schematic illustration of the preparation of 3D/2D perovskite thin films. 3D/2D perovskite thin films are prepared by depositing a hexylammonium bromide (HABr) precursor on the 3D bulk perovskite thin film, followed by thermal annealing for *in-situ* conversion. **b,** Cross-sectional bright field transmission electron microscopy (TEM) images of control, 10 mM HABr treated, and 50 mM HABr treated samples stacked with a hole transport layer (Spiro-OMeTAD), Au metal contact, and Pt layer that was deposited for TEM lamella preparation. The 2D thickness values reported below each image were determined from spatially calibrated line profiles (N=20 over 8 different images per sample). **c,** Time-resolved photoluminescence (TRPL) decay traces of control, 10 mM, and 50 mM HABr treated samples excited through the glass side. **d,** Intensity-dependent TRPL decay traces along with global fits (black lines) for a champion 50 mM HABr treated sample. **e,** Reported charge carrier lifetimes of direct bandgap semiconducting materials including perovskites, gallium arsenide (GaAs), indium phosphide (InP), and cadmium telluride (CdTe) over the past 50 years.

## Simultaneous Formation of Heterostructure and Bromine Gradient in 3D Layer

When the HABr concentration is tuned properly, we are able to set a new benchmark for the charge carrier lifetime in direct bandgap materials, yet it still remains unclear why these treatments are so effective at reducing non-radiative loss at surfaces and interfaces. To better understand how the HABr surface treatment impacts the structure and composition of the 3D surface, we directly visualize the compositional changes using a unique combination of nanoscale characterization techniques. Figure 2a and b show cross-sectional high-angle annular dark field (STEM-HAADF) images of a control and a 50 mM HABr treated sample. Figure 2c and d show STEM-EDX elemental maps of these same regions, where the control sample exhibits a uniform halide composition (see Figure S6 and S7 for maps of other elements) with an average Br/(Br+I) (from now on referred to as BBI, consistent with CIGS/GGI nomenclature)[21] of 0.006 ± 0.002, which matches well with the solution stoichiometry ($BBI_{solution}$ = 0.008). In contrast, the treated sample shows large variations in halide composition, where Br is primarily concentrated at the top surface and grain boundaries (see Figure S8 for maps of all elements), which are regions known to possess high defect densities.[4,22] We also note clear evidence of $PbI_2$ deposits at the grain boundaries in the control, which disappear after the HABr treatment (see Figures S6 and S7), indicating the *in-situ* surface reaction of HABr with $PbI_2$ to form a lower dimensional perovskite, consistent with other reports.[23] These *in-situ* reactions are strongly dependent on the as-grown perovskite surface chemistry, the location and concentration of $PbI_2$, and the size and chemical structure of the aromatic or aliphatic organic salt.[24]

From first inspection of Figure 2d, it is unclear whether the Br-rich strip is located in the 2D layer or the bulk 3D layer. As this would have major implications for the electronic structure of these materials and their shared interface, we perform a detailed analysis of key elemental line profiles of the sample stack to identify the interface positions and composition (see SI for fitting details and Figure S9). Using this analysis, Figure 2e,f show the positions of each interface along with fitting ranges for the control and 50 mM HABr treated samples. Importantly, Figure 2f shows that in the treated sample, the Br concentration peaks in the bulk 3D layer and a small amount of Br is present in the 2D layer. In fact, not only is Br concentrated in the 3D layer, but we are able to directly visualize a gradient extending >100 nm into the bulk film (see Figure S10 for additional line profiles). This gradient and the relative halide ratios are further confirmed through nanoscale depth-dependent time of flight secondary ion mass spectrometry (ToF-SIMS) measurements of samples treated with varying concentrations of HABr. Importantly, we show that not only is the thickness of the iodide-rich 2D layer tunable (see Figure 1b), but also the magnitude and depth of the BBI gradient. We note that the gradient appears stable under 1-Sun illumination and undergoes a small amount of halide redistribution when exposed to heat (i.e. 100°C – see Figure S11) where, importantly, other work has shown Br gradients lead to improved device stability.[25] Interestingly, we find that this gradient cannot be formed by just using a different bromide salt such as formamidinium bromide (FABr - see Figure S12) unlike in previous studies,[25] suggesting that HA is necessary to restructure the surface and facilitate the partial halide exchange.

Previously, halide compositional gradients have been pursued through surface treatments with FA and methylammonium halides,[25,26] but these treatments do not lead to the formation of a 2D layer, which is beneficial for device stability.[27] Here we show the simultaneous formation of a compositional gradient in the bulk layer as well as a 2D perovskite layer on top, which cooperatively provide both performance and stability enhancements.[3,6,27] Before now, each of these beneficial effects have been achieved separately through complex processing, but here we show both can be controllably formed through one simple deposition step.

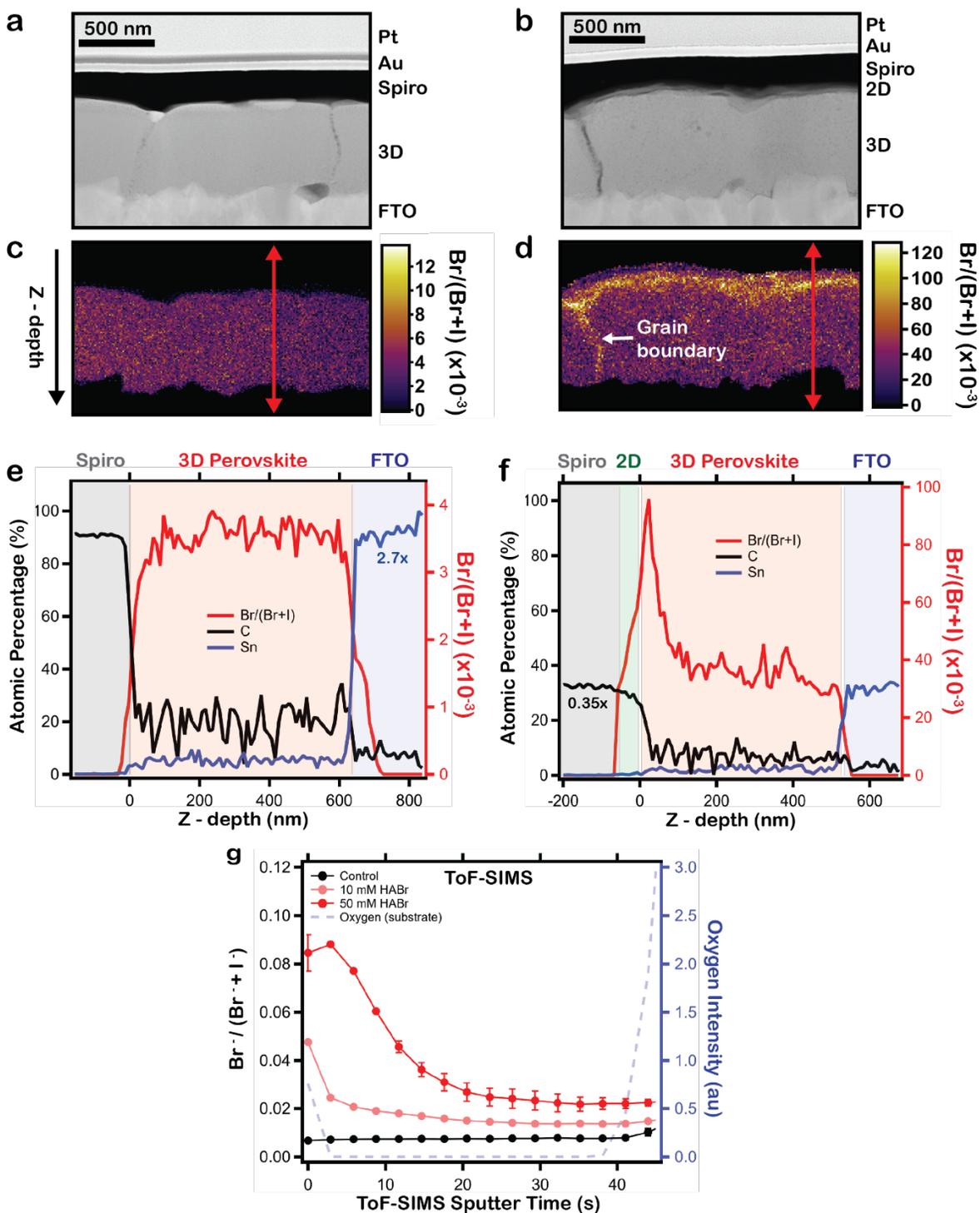

**Figure 2.** Cross-sectional STEM-HAADF images of the **a,** control and **b,** 50 mM HABr treated film. Energy dispersive X-ray spectroscopy maps showing the fraction of bromide (Br/(Br+I)) for the **c,** control and **d,** 50mM HABr treated films using pseudo-color scale bars. Atomic percentage line profiles for carbon (C), tin (Sn), and the bromide fraction for the **e,** control and **f,** 50 mM HABr treated film at the lines indicated by the red arrows in **c,** and **d**. Grey vertical lines denote

interface positions and ranges determined from fits. **g,** Depth-dependent time of flight secondary ion mass spectrometry (ToF-SIMS) showing the $Br^-$ / ($Br^-$ + $I^-$) ratio of control, 10 mM, and 50 mM HABr samples. Error bars are standard deviations for three depth-profiles measured on the same sample.

**Impact of HABr on Band Structure and Charge Carrier Dynamics**

With deeper insights into how the local composition changes as a function of film depth and HABr concentration, we next use PL spectroscopy and ultraviolet photoemission spectroscopy (UPS) to probe how the compositional gradient impacts the electronic structure and charge carrier distributions upon photoexcitation. As the gradient is primarily located at the top surface and within the first 100 nm, we chose to photoexcite the sample with a 405 nm laser in order to probe the dynamics closest to the surface. Figure 3a and b show spectral and time-resolved PL measured using a streak camera over a 5 ns time-window for both the control and 50 mM HABr treated samples. For the control sample, we only observe one emission peak, consistent with the 3D bulk emission. In contrast, for the 50 mM HABr treated sample, we observe several new peaks at higher energy. In particular, the most emissive peak at 570 nm, we identify as $n = 2$, $HA_2FAPb_2I_7$ (see Figure S13). This further confirms the EDX result that after HABr treatment, Br undergoes halide exchange in the 3D bulk and the 2D layer formed on top is iodide rich, which is consistent with Figure 2 and recent observations by Liu *et al.*.[28] We use a multi-peak fitting analysis to extract key information about the evolution of each peak and show the center emission for the 3D perovskite when treated with 10 and 50 mM HABr (white overlay in Figure 3a,b). We observe a concentration dependent redshift in the 3D emission profile for the HABr treated samples as a function of time (Figure 3c) and a smaller shift for the control sample. These trends in the center emission energy shift as a function of HABr concentration match well with the predicted bandgap gradient extending through the sample thickness (Figure S14). We note that these shifts in energy are not an artifact of the experimental setup (see Figure S15) and do not appear when excitation was performed from the glass side (i.e. where the 2D layer is absent - see Figure S16).

The depth-dependent nanoscale compositional measurements in Figure 2 along with the redshift in the emission energy are strong evidence for the formation of bandgap gradings that extend from the top surface and grain boundaries into the bulk of the material. Next, we perform depth-dependent UPS measurements to further understand how HABr modifies the electronic band structure. Figure 3d shows the valence band maximum (VBM) as a function of depth and HABr concentration. The magnitude of band bending at the surface increases with HABr and Br concentration (see Figure S17 for full spectra), consistent with previously reported UPS measurements.[24] Figure 3e summarizes how treatment with HABr locally changes the composition and electronic structure by forming an iodide-rich 2D layer and Br gradient at the top surface and grain boundaries. This results in a wider bandgap at defective surfaces and larger overall band bending that extends deep into the sample. Excitingly, the compositional gradients and surface energetics can be precisely tuned using 2D/3D passivation strategies. This mechanism of forming surface fields is vastly different than other PV technologies that rely on creating compositional gradings during growth by varying the gas composition, such as in GaAs[12] and CIGS[29], or by utilizing the diffusion from an alloyed metal contact (i.e. Al-back surface field in Si).[30]

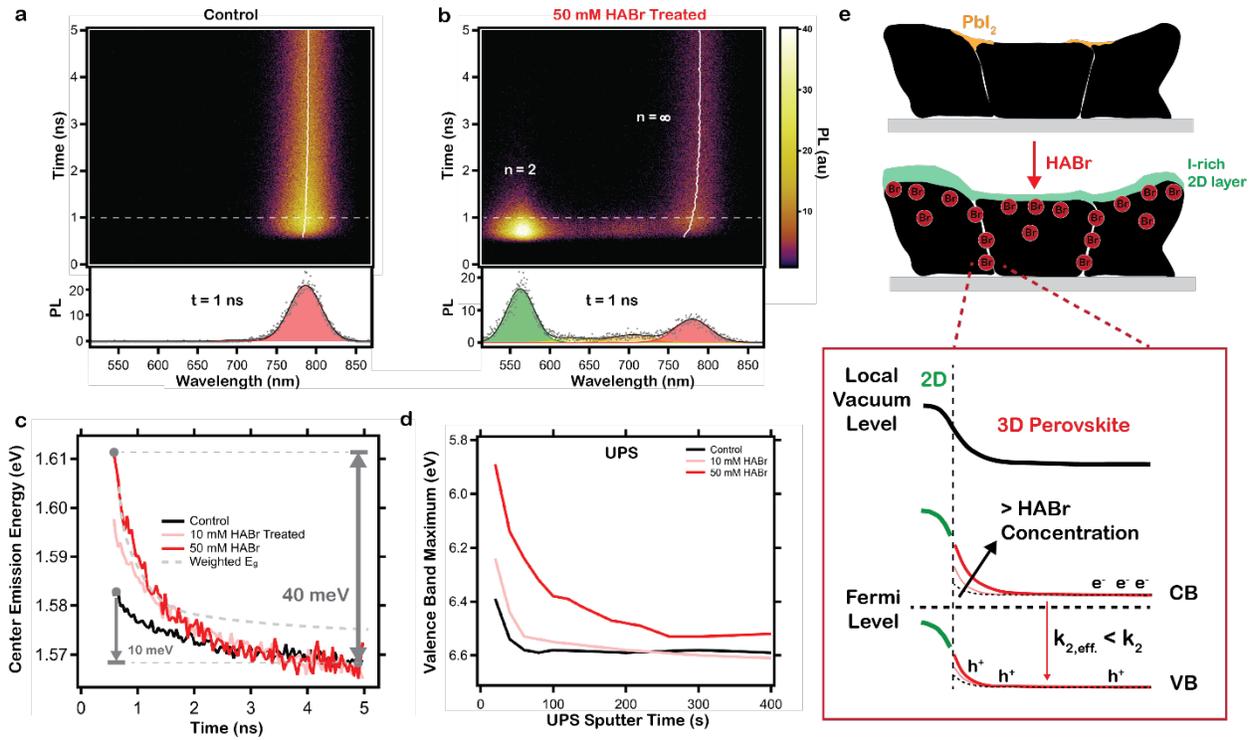

**Figure 3.** Photoluminescence spectra as a function of time along with the center emission energy (overlaid white line) for the **a,** control and **b,** 50 mM HABr treated films when excited from the top surface. An example of the PL spectra and Gaussian fits at $t$ = 1 ns (dashed white line) are shown below the streak camera data. **a,** and **b,** share the same pseudo-color scale bar. **c,** Center emission energy as a function of time for the control (black line), 10 mM, and 50 mM HABr treated (red lines) samples along with the predicted change in the bandgap ($E_g$, dashed blue line) as a function of time and based on the numerical simulation in Figure S14. **d** Valence band maximum determined from depth-dependent ultraviolet photoemission spectroscopy (UPS) of a control, 10 mM hexylammonium bromide (HABr), and 50 mM HABr treated samples. **e** Schematic illustration of the working mechanism of HABr treatments and their impact on the electronic energy structure as a function of concentration (not to scale). Electrons ($e^-$) in the conduction band (CB) and holes ($h^+$) in the valence band (VB) are spatially separated *via* the surface fields.

**Ultralow Interfacial Recombination at HTL**

The precise control of bandgap gradings and surface fields are design principles that have led to world-record PCEs in a host of PV materials, including CdTe,[31] GaAs,[32] CIGS,[29,33] quantum dots,[34] and tandem architectures.[12,35,36] Contrarily, the ability to engineer surfaces and interfaces, especially with 2D/3D heterostructures, is far less developed for perovskites.[2] We note that the built-in electric field due to the heterostructure and bandgap grading is dependent both on the grading's depth and magnitude, with common values for other PV technologies ranging from 1.5 to 10 kV cm$^{-1}$.[34,37] In this work, we measure the energy grading to be at least 50 meV, with additional band bending of 100's of meV due to the 2D layer (see Figure 3d). This appears comparable to most CIGS bandgap gradings (~200 meV),[37] but here the grading occurs over such a shallow depth that it leads to a similar or even greater built-in electric field of ~ 9 kV cm$^{-1}$ (Figure 4a). Importantly, Figure 4a shows the impact of a 50 meV energy grading on the charge carrier density distributions (i.e. PL $\propto$ k$_2$np - see Figure S18 and S19 for electron and hole density maps) as a function of depth and time when excited from the substrate side (opposite of 2D/3D interface). At early times, emission is concentrated under the generation profile and then spreads through the film thickness due to diffusion. The ~ 9 kV cm$^{-1}$ field strength leads to a ~80% reduction in the electron density near the surface/interface, which are known to possess high densities of defects in perovskites.[38] Surface fields not only repel carriers from defective regions at the surface, but also cause spatial separation of electrons and holes which has previously been shown to lead to depressed radiative recombination rates (i.e. lower PLQE) and slowed recombination in materials such as InP.[39,40] We capture this effect in the lower fitted $k_{2,eff}^{int}$ in our samples (Figures 1d and 3e). In fact, we find that slowed recombination is a general phenomenon in other 2D/3D heterostructures (i.e. treatment with phenethylammonium iodide (PEAI), see Figure S20), and can be explained through band bending (see SI and Figure S21), where the extent of field induced charge separation can be evaluated from the $k_{2,eff}^{int}$ value along with the appearance of a drift-induced fast PL decay component[41] when excited from the top surface (see SI and Figures S22 and S23). PL spectroscopy therefore serves as a simple and readily accessible tool to quickly evaluate surface energetics[42] and tune new chemical surface modifications and heterostructures.

Next, we study device performance as a function of HABr concentration using our typical high performing device architecture (see SI). We find that 10 mM HABr leads to much better performance than the 50 mM HABr device (Figure S24), likely a result of better band alignment and optimized 2D layer thickness. In order to better understand the impact of both the 2D layer and the bandgap gradient on interfacial recombination and device performance, we focus on the 10 mM treatment after depositing a hole transport layer (HTL), Spiro-OMeTAD, on top of the 2D/3D heterostructure. Figure 4b shows the time-resolved PL decay trace of the control/HTL bilayer decaying several orders of magnitude over the first 100s of nanoseconds, consistent with other reports.[43] Conversely, the treated/HTL sample has a fast initial decay, which is attributed to hole extraction (i.e. majority surface recombination velocity (SRV),[43] followed by a very slow decay (i.e. minority SRV). We quantify these differences in decay kinetics by determining the upper limit of the SRV, following the approach described by Wang *et al.* (see SI).[43,44] For the control sample, we fit an SRV of < 3500 cm/s, consistent with the 3100 cm/s value previously reported for Spiro-OMeTAD.[43] The treated sample demonstrates an SRV orders of magnitude lower, reaching values as low as < 7 cm/s. To the best of our knowledge, this is the lowest SRV reported for any perovskite interfaced with a charge transport layer,[43,45] and is approaching values of passivated perovskite single crystals with no HTL (see Figure S25).[46] Figure 4d highlights this trend, where the control PLQE and implied $V_{oc}$ ($V_{OC}^{imp.}$- see SI for details) drop significantly after introduction of Spiro-OMeTAD, which is commonly observed for perovskites interfaced with any charge transport layer.[8,47] Contrarily, the treated sample PLQE and $V_{OC}^{imp.}$ does not decrease and may even slightly improve. To place this value in context, a ~6% PLQE for a perovskite interfaced with a transport layer is amongst the highest reported to date.[24] We confirm the quality of this 2D layer by fabricating n-i-p solar cells following our previous studies,[3] showing that the introduction of this 2D layer improves device PCE from ~21% to values as high as 24% and up to 25% with an antireflective (AR) coating (Figure 4d), values that have previously been certified.[3] Importantly, key enhancements are achieved in the open circuit voltage ($V_{oc}$) (92% of radiative limit – see SI and Figure S26) and fill factor (FF), indicating a significant reduction in non-radiative loss[48] (see refs[3,6] for device statistics). Importantly, the $V_{OC}^{imp.}$ values in Figure 4c are consistent with the control and treated device $V_{oc}$'s (see Figure 4d) and highlight the ultralow non-radiative loss observed in the treated devices (Figure S26).[3] We note that similar treatments can likely be optimized in p-i-n architectures as previously shown with guanidinium bromide (GABr).[49]

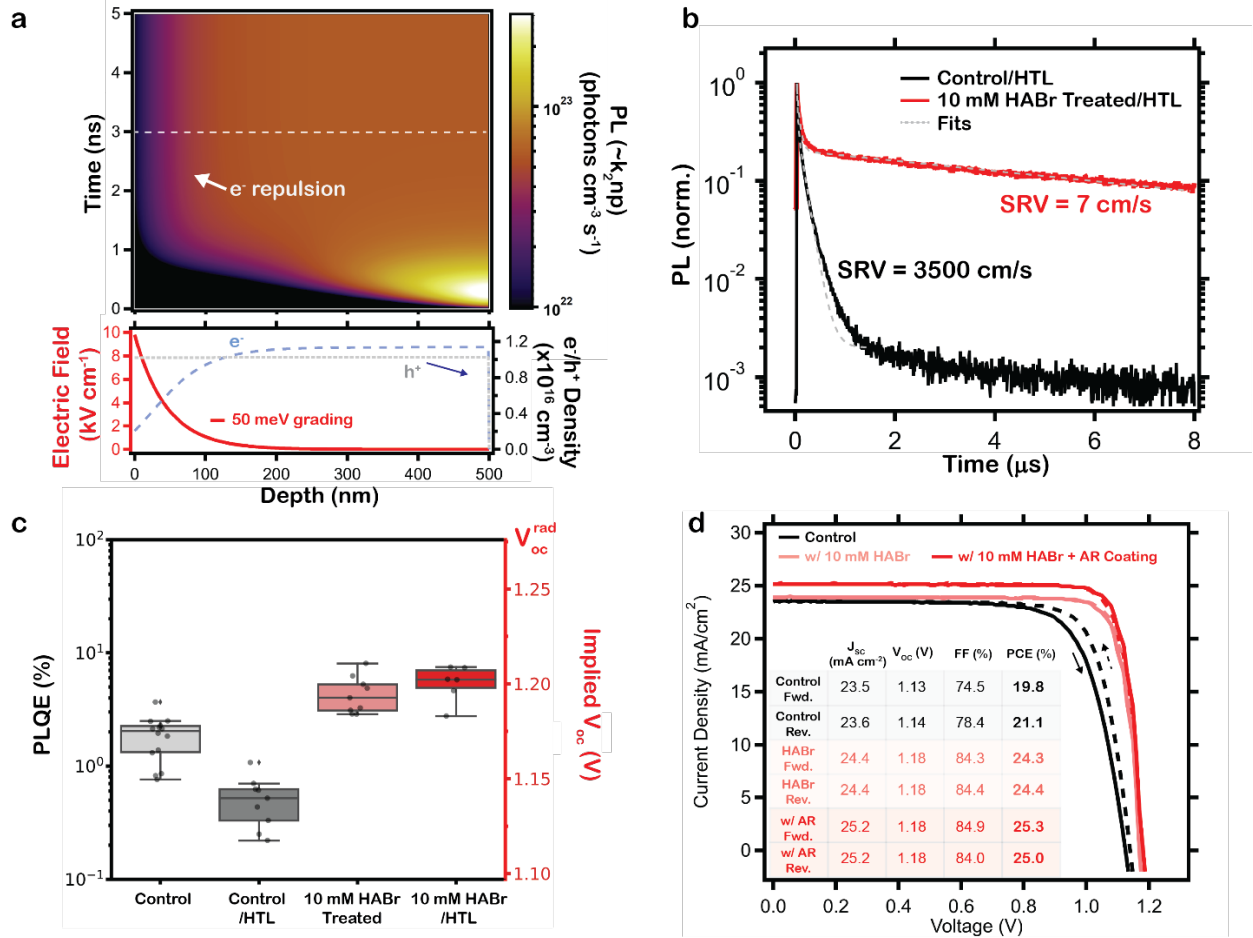

**Figure 4. a,** Simulated PL map as a function of time through the film thickness with a 50 meV energy grading. Electron and hole density line profiles in the subfigure are taken at t = 3 ns (white dashed line). **b,** Measured time-resolve photoluminescence (PL) decay traces of a control 3D versus a hexylammonium bromide (HABr - 10 mM) treated sample when interfaced with the hole transport layer (HTL) Spiro-OMeTAD along with biexponential fits (dotted gray lines). **c,** Box and whisker plots of measured PL quantum efficiency (PLQE) and calculations of implied open circuit voltages ($V_{oc}$) of a control (sample size, $n$ =14) and control/hole transport layer (HTL) ($n$ = 9) versus a HABr treated (10 mM) film ($n$ = 9) and treated/HTL bilayer ($n$ = 6). All measurements were taken on distinct samples, with HTL layers deposited on previously measured control and treated samples. Center line, median; box limits, upper and lower quartile; whiskers, 1.5x interquartile range; diamond data points, outliers. **d,** Current density-voltage curves of the forward (solid lines) and reverse (dotted lines) scans of a control (black), 10 mM HABr treated (red), and 10 mM HABr treated perovskite photovoltaic device with an antireflective (AR) coating.

Although the use of 2D/3D heterostructures have become nearly ubiquitous in achieving record-setting PCE's for perovskites,[3,50] it has been unclear how these types of chemical treatments lead to improved charge carrier dynamics on the nanoscale.[2] For the first time, we reveal the key features a 2D/3D interface that enabled HABr-treated perovskites to reach the >25% PCE milestone and have identified selection rules and design principles that could be further exploited to achieve charge carrier lifetimes and interfacial recombination velocities necessary to reach 30% PCE.[45] We show that HABr treatment primarily forms an iodine-rich 2D layer on top of the 3D bulk layer and Br penetrates >100 nm into the 3D layer and selectively targets the top surface and grain boundaries. The unique formation of these surface electric fields is different than any other PV technology to date and therefore establishes new directions in the control of local precursor (i.e. $PbI_2$) concentration at the surface, the initial stoichiometry of the bulk perovskite and halide composition, as well as the concentration and structure of the bulky A-site cation. Importantly, we have identified key features of PL lifetime measurements that can be used as a rapid feedback tool to optimize surface fields and carrier distributions. The fundamental insights in this work naturally transfer over to other fields where simultaneous formation of heterostructures and gradients could be beneficial, for example in the formation of highly efficient core-shell quantum dot structures for LEDs and other optoelectronic devices.

## Data Availability

The data that support the findings of this study are available from the corresponding authors upon reasonable request.

## Code Availability

## Acknowledgements

D.W.D., R.B., M.L., M.B. and V.B. acknowledge support for this project through the MIT-Tata GridEdge Solar Research Program, which is funded by the Tata Trusts. J.J.Y. was funded by the Institute for Soldier Nanotechnology (ISN) grant W911NF-13-D-0001. This work has also been supported in part by the Department of Energy (DOE), Office of Energy Efficiency and Renewable Energy (EERE), Award Number: DE-EE0009512. M.L. and R.B. acknowledge support from the National Science Foundation Graduate Research Fellowship under Grant Number: 1122374. R.B. acknowledges support from MathWorks through the MathWorks Engineering Fellowship. F.U.K. thanks the Jardine Foundation and Cambridge Trust for a doctoral scholarship. S.S.S. was supported by a grant from the Korea Research Institute of Chemical Technology (KRICT), South Korea (KS2022-10); the Korea Institute of Energy Technology Evaluation and Planning (KETEP) and the Ministry of Trade Industry & Energy (MOTIE) of the Republic of Korea (NO. 20183010014470). Part of this work was conducted at the Molecular Analysis Facility, a National Nanotechnology Coordinated Infrastructure (NNCI) site at the University of Washington, which is supported in part by funds from the National Science Foundation (awards NNCI-2025489, NNCI-1542101), the Molecular Engineering & Sciences Institute, and the Clean Energy Institute. We would also like to acknowledge KARA (KAIST Analysis center for Research Advancement) for help in conducting UPS measurements. D.W.D. thanks David Fenning (UCSD), Roderick MacKenzie (U. Nottingham), Ana Kanevce (NREL), Hannah Smith (Princeton), Sam Stranks (U.

Cambridge), Thomas Kircharts (Forschungszentrum Jülich), and Lisa Krückemeier for valuable discussions and sharing literature survey data of perovskite carrier lifetimes.

**Contributions**

D.W.D., J.J.Y., M.B., and V.B. conceived and designed the experiments. D.W.D., J.J.Y., M.L., R.B., and B.D.D. performed the optical characterization of the perovskite films. F.U.K. performed the (S)TEM/EDX measurements and D.W.D. and F.U.K. performed the analysis of the data with supervision from C.D.. D.W.D., R.B., and M.L. wrote the MATLAB and Python code for fitting PL data and performing drift-diffusion simulations. J.J.Y. prepared the perovskite films and devices with supervision from S.S.S.. D.J.G. and K.H. performed ToF-SIMS measurements and D.W.D and D.J.G. performed analysis of the data. D.W.D wrote the first draft of the manuscript with early drafts edited by J.J.Y. and all authors contributed feedback and comments. M.G.B and V.B. directed and supervised the research.

**Competing Interests**

V.B. is an advisor to Swift Solar, a US company developing perovskite photovoltaics, and is the co-founder of Ubiquitous Energy, a US company developing visibly transparent photovoltaics. D.W.D. is a co-founder of Optigon, Inc., a US company developing metrology tools for the photovoltaics industry.

**Inclusion & Ethics**

The contributions of all researchers involved in the design, execution, and reporting of this study were carefully evaluated for authorship criteria. Contributors who did not meet all criteria for authorship are listed in the Acknowledgements section. The roles and responsibilities amongst all collaborators were agreed upon ahead of the execution of the research.

**Additional Information**

**Supplementary Information**

This file contains Supplementary Methods, Supplementary Figures S1-S19, and Supplementary References.

**Materials & Correspondence**

Correspondence and requests for materials should be addressed to Vladimir Bulović.

# Supporting Information for:

# Reduced Recombination *via* Tunable Surface Fields in Perovskite Solar Cells

Dane W. deQuilettes,[1] Jason Jungwan Yoo,[2,3] Roberto Brenes,[4] Felix Utama Kosasih,[5] Madeleine Laitz,[4] Benjia Dak Dou,[1] Daniel J. Graham,[6] Kevin Ho,[7] Seong Sik Shin,[3*] Caterina Ducati,[5] Moungi Bawendi,[2*] Vladimir Bulović[1,4*]

[1] Research Laboratory of Electronics, Massachusetts Institute of Technology, 77 Massachusetts Avenue, Cambridge, Massachusetts, USA

[2] Department of Chemistry, Massachusetts Institute of Technology, 77 Massachusetts Avenue, Cambridge, Massachusetts, USA

[3] Division of Advanced Materials, Korea Research Institute of Chemical Technology, Daejeon, Republic of Korea.

[4] Department of Electrical Engineering and Computer Science, Massachusetts Institute of Technology, Cambridge, MA, USA.

[5] Department of Materials Science and Metallurgy, University of Cambridge, 27 Charles Babbage Road, Cambridge CB3 0FS, UK.

[6] Department of Bioengineering, University of Washington, Box 351653, Seattle, Washington, USA

[7] Department of Chemistry, University of Washington, Box 351700, Seattle, Washington, USA

**\*Corresponding Authors:** bulovic@mit.edu, mgb@mit.edu, and sss85@krict.re.kr

## Sample Preparation

### Chemicals

DI water, urea, hydrochloric acid (HCl, 37 wt. % in water), thioglycolic acid (TGA, 98%), SnCl$_2$·2H2O (>99.995%), dimethylformamide (DMF), dimethyl sulfoxide (DMSO), diethyl ether, chlorobenzene, chloroform, isopropyl alcohol, lithium Bis(trifluoromethanesulfonyl)imide salt (Li-TFSI), and 4-tert-butylpyridine (tBP) were purchased from Sigma-Aldrich. 2,2',7,7'-Tetrakis(N,N -di-p -methoxyphenylamino)-9,9'-spirobifluorene (Spiro-OMeTAD, LT-S922) and Tris(2-(1H -pyrazol-1-yl)-4-tert-butylpyridine)-cobalt(III)Tris(bis(trifluoromethylsulfonyl)imide)) salt (Co(III) TFSI) were purchased from Lumtec. Methylammonium chloride (MACl), formamidinium iodide (FAI), methylammonium bromide (MABr), and n-hexylammonium bromide were purchased from GreatCell Solar Materials. Lead iodide (PbI$_2$) and lead bromide (PbBr$_2$) were purchased from TCI America. Au pellets were purchased from Kurt J. Lesker.

## Perovskite Thin Film and Device Fabrication

The perovskite samples were prepared following our previous reports.[1] Briefly, the perovskite solution, which was prepared by mixing 1.53 M PbI$_2$, 1.4 M FAI, 0.5 M MACl, and 0.0122 M MAPbBr$_3$ in DMF:DMSO=8:1, is deposited onto either a glass substrate or a SnO$_2$ electron transport layer via spin coating at 1000 rpm for 10 sec, and 5000 rpm for 30 sec (both 2000 rpm ramp), followed by dripping of 600 uL of diethyl ether 10 sec into the 5000 rpm setting. We note that an aqueous solution of potassium iodide (KI) is spin coated on top of the SnO$_2$ layer prior to perovskite deposition. This KI treatment is done in all films and/or devices and therefore is not unique to devices that are treated with HABr. Next, the perovskite film is annealed at 100 °C for 60 min. For the 2D perovskite passivation, hexylammonium bromide dissolved in chloroform was deposited at 3000 rpm for 30 sec and annealed at 100 °C for 5 min. The hole transporting layer (HTL) was deposited by mixing 50 mg of Spiro-OMeTAD, 19.5 µL of tBP, 5 µL of Co(III) TFSI solution (0.25 M in acetonitrile), 11.5 µL of Li-TFSI solution (1.8 M in acetonitrile), and 547 µL of chlorobenzene. 70 µL of the HTL solution was loaded onto the perovskite substrate and spin coated at 4000 rpm for 20 sec (2000 rpm ramp). The Au electrode (100 nm) was deposited by thermal evaporation.

## Device Characterization

Current density-voltage (J-V) curves were recorded using a solar simulator (Newport, Oriel Class AAA, 91195A) and a source meter (Keithley 2420) in ambient lab environment. The illumination was set to AM 1.5G and calibrated to 100 mW/cm$^2$ using a calibrated silicon reference cell. The step voltage was 10 mV and the delay time was 50 ms. The active area was controlled by using a dark mask with a defined aperture of 0.096 cm$^2$.

## (Scanning) Transmission Electron Microscopy ((S)TEM) Methods

For (S)TEM characterization, cross-sectional lamellae were prepared with an FEI Helios Nanolab Dualbeam FIB/SEM following a standard procedure described elsewhere.[2] The lamellae were immediately transferred into an FEI Tecnai Osiris (S)TEM, minimizing air exposure to ~2 min. This instrument was operated with a 200 kV beam. Bright field TEM images were acquired using a Gatan UltraScan1000XP camera, with a pixel size of 1.2 nm and an electron dose of ~56 e⁻/Å². This dose is approximately half of the reported damage threshold for hybrid halide perovskites.[3] STEM-HAADF images were acquired using a Fischione detector, with a beam current of ~140 pA and a dwell time of 1 µs/pixel. STEM-EDX data was obtained using a Bruker Super-X silicon drift detector system with a collection solid angle of ~0.9 sr, a beam current of ~140 pA, a dwell time of 30 ms/pixel, a spatial sampling of 10 nm/pixel, and a spectral resolution of 5 eV/channel. The electron dose for STEM-EDX was ~2620 e⁻/Å², a value previously optimised with respect to beam-induced specimen damage and EDX data quality.[2] STEM-EDX data was processed in HyperSpy, an open-source Python package for multidimensional data analysis.[4] First, the EDX data was spectrally rebinned to 20 eV/channel, then denoised using principal component analysis to increase the signal-to-noise ratio.[5,6] Subsequently, the background-corrected intensities of X-ray peaks of interest were extracted. To obtain quantitative elemental maps, Cliff-Lorimer quantification was performed in each pixel using the X-ray peak intensity values.[7] Bright field TEM images and thickness determination measurements were performed using the open-source software ImageJ (https://imagej.nih.gov/ij/). Analysis of STEM-EDX images was performed in Igor Pro 7 using the Image Processing Package. Ratio maps are produced by dividing the quantitative data matrix of one element by the other. Thresholding masks were applied to I and Br/(Br+I) maps as small background counts could make the Br/(Br+I) ratio values unphysically high in the region above the perovskite layer (i.e. Spiro-OMeTAD layer). The thresholding value was set right before the point at which I background counts started to appear at the top of the Spiro layer and far away from the regions where I counts are expected (i.e. in the 2D and 3D perovskite layers).

## Time of Flight-Secondary Ion Mass Spectrometry (ToF-SIMS)

ToF-SIMS depth profiles were acquired on a IONTOF TOF.SIMS5 spectrometer using a 25 keV $Bi^{3+}$ cluster ion source in the pulsed mode. Depth profiles were carried out in the non-interlaced mode. This mode alternates between collecting data and then sputtering the surface to remove a layer of material. Each data acquisition is considered a 'layer' in the profile. Data was acquired in the negative ion mode using the delayed extraction mode over a mass range of m/z = 0 to 800. The primary ion beam was rastered over a 100 micron x 100 micron area using 256 pixels x 256 pixels. The data acquisition location was centered within the sputter crater to avoid any artifacts from the crater edges. The ion source was operated with at a current of 0.06 pA. Secondary ions of a given polarity were extracted and detected using a reflectron time-of-flight mass analyzer. The primary ion dose per layer was 2.5 x $10^{11}$ ions/cm². Sputtering was carried out using an argon gas cluster ion beam using argon 1000 clusters at 20 keV. The sputter beam was rastered over a 500 micron x 500 micron area using a dose per layer of 5 x $10^{13}$ ions/cm² . Since the sputter rate is not known for these samples, the sputter depth cannot be determined. Negative ion spectra were calibrated using the $C_2H^-$, $PbI_2^-$ and $PbI_3^-$ peaks. Calibration errors were kept below 25 ppm.

## Ultraviolet Photoemission Spectroscopy (UPS)

The perovskite thin film samples are loaded into an ultrahigh vacuum chamber (10-7-10-8 Torr) for the depth profile UPS measurements (Axis Supra, Kratos). A UV source (He I 21.2 eV) with the beam size of ~8 mm and a pass energy of 10 eV with a bias of -9V, dwell time of 100ms, and an energy step size of 0.025 eV. The UPS spectra is shown in binding energy that was calibrated using a vacuum deposited Au. The secondary electron cutoff is used to determine the vacuum level with respect to the Fermi level and the valence band maximum is calculated by determining the valence band edge onset in a log plot. For etching, Ar cluster etching was used with ion energy of 10 keV with the raster size of 4 mm.

## Time-Resolved Photoluminescence Measurements

A 405 nm or 470 nm pulsed diode laser (LDH series) was used to photoexcite samples with repetition rates ranging from 5-125 kHz. The sample PL emission was filtered through a 700 nm long pass filter and directed to either a Micro Photon Devices (MPD) PDM Series single photon avalanche photodiode with a 50 $\mu$m active area or a PMA Hybrid 50 detector. Photon arrival times were time-tagged using a time-correlated single photon counter (Picoquant- TimeHarp 260) and data was collected using the Picoquant TimeHarp 260 software.

## Photoluminescence Quantum Efficiency (PLQE) Measurements

PLQE measurements were acquired using a center-mount integrating sphere setup (Labsphere CSTM-QEIS-060-SF) and Ocean Optics USB-4000 spectrometer. The integrating sphere setup was intensity calibrated with a quartz tungsten halogen lamp (Newport 63355) with known spectral irradiance. A fiber-coupled 405 nm diode laser in continuous-wave (CW) mode (PDL-800 LDH-P-C-405B) was collimated with a triplet collimator (Thorlabs TC18FC-405) to produce a beam with an approximate $1/e^2$ diameter of 900 $\mu$m (measured with a CCD Camera beam profiler, Thorlabs BC106N-VIS/M). The laser power density was set to ~70 mW/cm$^2$ in order to create an absorbed photon flux roughly equivalent to AM1.5 solar illumination conditions.[8] Data was collected and analyzed using custom software written in Python. The PLQE was determined by following the protocol described by de Mello *et al.*,[9] with a scattering correction.

## Streak Camera Spectroscopy Measurements

Optical spectroscopy was performed using a Nikon Eclipse-Ti inverted microscope fitted with an infinity corrected 20 × dry objective (Nikon S Plan Fluor, NA = 0.45). A 405 nm pulsed diode laser (PDL-800 LDH-P-C-405B, 300 ps pulse width) was used for excitation with repetition rate of 62.5 kHz. The laser beam and sample emission were filtered through a 405 nm dichroic beamsplitter (Nikon DiO1-R405) and 450 nm long pass filter then coupled in free space into a streak camera (Hamamatsu C5680) equipped with a slow speed sweep unit (M5677). The time delay between the laser source and sweep unit was controlled using a digital delay generator (Stanford Research Systems, Inc. Model DG645). Data was collected using the time-correlated single photon counting mode in the HPD-TA 8.4.0 software (Hamamatsu). Individual spectral peak information was extracted by fitting each temporal slice with either a single Gaussian function (for control samples) or the summation of four Gaussian functions in the case of the 2D/3D heterostructure using custom code written in Python.

**Interface and Edge Detection Using STEM-EDX Line Profiles**

STEM-EDX line profiles in regions of interest were fit with a Gaussian error function (i.e. integral of Gaussian function) of the form:

$$f(d) = \frac{A}{2}\left[1 + erf\left(\frac{d-\mu}{\sqrt{2}\sigma}\right)\right] + C \tag{S1}$$

where $A$ is the amplitude, $d$ is the depth, $\mu$ is the interface position between the two layers, $\sigma$ is the Gaussian width, and $C$ is the offset.

The variables $A$, $\mu$, $\sigma$, and $C$ were set as free variables and a least squares cost/objective function was minimized using a truncated Newton algorithm (i.e. New Conjugate-Gradient Method) as implemented in the SciPy library in Python.

**Global Fitting of Intensity Dependent Time-Resolved PL**

The time-resolved charge carrier recombination kinetics have shown to following a standard rate equation over a wide range of carrier densities[10-12] of the form:

$$\frac{dn}{dt} = -k_1 n - k_2 n^2 - k_3 n^3 \tag{S2}$$

where $n$ is the electron charge carrier density, $k_1$ is the first-order Shockley-Read-Hall (SRH) trapping (non-radiative) rate constant, $k_2$ is the (effective) internal second-order band-to-band (radiative) recombination rate constant, and $k_3$ is the third-order Auger (non-radiative) recombination rate constant.

Solutions to this ordinary differential equation, $n$, were numerically solved using the SciPy library in Python. The simulated PL decay traces are calculated using the $n$ values as inputs into equation S3:

$$PL(t) = P_{esc} k_2 n(t)^2 \tag{S3}$$

The measured PL signal, y(t), is then scaled with a scaling factor, A, and an offset, C

$$y(t) = A \cdot PL(t) + C \tag{S4}$$

Above, $P_{esc}$ is the escape probability calculated following previous reports.[8]

$$P_{esc} = \frac{\int_0^\infty a(E)\phi_{BB}(T,E)dE}{\int_0^\infty 4\alpha(E)dn_r^2(E)\phi_{BB}(T,E)dE} \qquad (S5)$$

Here $a$ is the material absorptivity, $\alpha$ is the absorption coefficient, $d$ is the film thickness, $n_r$ is the refractive index, and $\Phi_{BB}$ is the black body spectrum given by:

$$\phi_{BB}(T,E) = \frac{2\pi E^2}{h^3 c^2} \frac{1}{e^{\left(\frac{E}{KT}\right)} - 1} \qquad (S6)$$

Initial carrier densities used as initial conditions were determined by measuring the laser power along with the beam spot size at the sample plane (with a beam profiler) and calculated using the following equation:

$$n_0 = \frac{P\left(1 - 10^{-OD}\right)\lambda_{exc}}{fhc(\pi r^2 d)} \qquad (S7)$$

Where $P$ is the laser power, OD is the optical density at the excitation wavelength, $\lambda_{exc}$ is the laser excitation wavelength, $f$ is the laser frequency/repetition rate, $h$ is the Planck constant, $c$ is the speed of light, $r$ is the $1/e^2$ radius of the Gaussian excitation profile, and $d$ is the film thickness.

At the highest laser power in our experiments, the maximum carrier density was calculated to be $\sim 1 \times 10^{16}\,cm^{-3}$ and therefore we ignored the Auger (i.e. $k_3 n^3$ term in equation S2) as it was expected to account for <1% of the total decay rate at t = 0 (where the carrier density would be highest).

Reduced chi-squared surfaces were generated by calculating the magnitude of the error vector (i.e. for three different decay traces) over a matrix of $k_1$ and $k_2$ values, logarithmically spanning 3-4 orders of magnitude. The minimum of this surface was determined through an automated search function as implemented in the NumPy library in Python.

**Determination of Confidence Intervals**

Custom Python code was written to perform support plane error analysis and used to determine the 95% confidence intervals using the methodology described in the PicoQuant Fluofit Manual (see section 6.3.1- Support Plane Method).[13] Here, the tolerance of the reduced chi-squared value, $\chi_{tol.}^2$, was determined using the equation:

$$\frac{\chi_{tol.}^2}{\chi_{min}^2} = 1 + \frac{p}{v}F(p,v,P) \qquad (S8)$$

The confidence interval surfaces are determined by defining a tolerance over which the error becomes unacceptable. The tolerance level is determined from F-statistics, where $F(p,v,P)$ takes into account the number of parameters, $p$, the degrees of freedom, $v$, and a probability P determined by the confidence interval of interest (i.e. 95%). For our simulation and fitting, $F(2,8,0.95)$, and therefore $\frac{\chi_{tol.}^2}{\chi_{min}^2} = 1.96$.

**Drift-Diffusion Numerical Simulations**

One-dimensional drift-diffusion simulations were performed using modified MATLAB code developed by Weiss *et al.*,[14] where the accuracy was previously verified against standard Sentaurus Technology Computer Aided Design (TCAD) simulations. Briefly, the separate electron and hole densities are numerically solved using a set of coupled partial differential equations (PDE's) in the MATLAB pdepe solver. The transport equations for electron ($n$) and holes ($p$) can be described by:

$$\frac{\partial n}{\partial t} = G + k_B T \mu_n \frac{\partial^2 n}{\partial z^2} - \mu_n \frac{\partial}{\partial z}(nE_n) + \sum_i (e_{e,i} - e_{c,i}) - R_{rad} \tag{S9}$$

$$\frac{\partial p}{\partial t} = G + k_B T \mu_p \frac{\partial^2 p}{\partial z^2} - \mu_p \frac{\partial}{\partial z}(pE_p) + \sum_i (h_{e,i} - h_{c,i}) - R_{rad} \tag{S10}$$

$k_B$ is the Boltzmann constant, $T$ is the temperature, $\mu_n$ and $\mu_p$ are the electron and hole mobilities, respectively, $z$ is the depth in the thin semiconducting film, $E_n$ and $E_p$ are the electric field magnitudes (i.e. strength), $e_{e,i}$, $e_{c,i}$, $h_{e,i}$, $h_{c,i}$ are the electron and hole emission and capture rates into a midgap defect state with index, $i$, and $R_{rad}$ is the total radiative recombination rate. These equations do not explicitly solve for the electric field profile through the use of the Poisson equation. Instead, the bandgap grading is used to calculate the electric field profile as shown in equation S17.

The emission and capture rates are calculated using standard textbook equations defined as:

$$e_{e,i} = n_{t,i}\sigma_{n,i}v_t N_C \exp\left(-\frac{E_{t,i}}{k_B T}\right) \tag{S11}$$

$$e_{c,i} = n(N_{t,i} - n_{t,i})\sigma_{n,i}v_t \tag{S12}$$

$$h_{c,i} = (N_{t,i} - n_{t,i})\sigma_{p,i}v_t N_V \exp\left(-\frac{E_g - E_{t,i}}{k_B T}\right) \tag{S13}$$

$$e_{c,i} = pn_{t,i}\sigma_{p,i}v_t \tag{S14}$$

Where $N_{t,i}$ is the total defect density, $E_{t,i}$ is the energy of the trap state relative to the conduction band energy, $\sigma_{n,i}$ and $\sigma_{p,i}$ are the electron and hole capture cross-sections, respectively, $v_t$ is the thermal velocity of the charge carriers, and $N_C$ and $N_V$ are the conduction and valence band effective density of states. The rate equation for the density of occupied trap states, $n_{t,i}$, can then be written as

$$\frac{\partial n_{t,i}}{\partial t} = -e_{e,i} + e_{c,i} + h_{e,i} - h_{c,i} \tag{S15}$$

The total radiative recombination rate is defined as

$$R_{rad} = k_2\big((n_0 + \Delta n)(p_0 + \Delta p) - n_0 p_0\big) \tag{S16}$$

*Electric field Strength Calculation*

As the Br/(Br+I) gradient measured in the main article is primarily expected to impact the conduction band energy ($E_C$),[15] the electric field strength for electrons is calculated by taking the derivative of the conduction band grading with respect to position. This same equation can also be modified to calculate the electric field profile for the valence band energy, which was done for Figure S21.

$$E_n = \frac{dE_C}{dz} \qquad (S17)$$

*Initial Conditions*

In equations S9 and S10, $G$ is the generation rate defined as $G(z,t) = \alpha_\lambda N_0 I_{laser}(\text{t})\exp(-\alpha_\lambda z)$. Where $\alpha_\lambda$ is the absorption coefficient ($2\text{x}10^5$ cm$^{-1}$) at the laser excitation wavelength (405 nm), $N_0$ is the absorbed photon flux determined from the excitation power and spot size, and $I_{laser}(t)$ is the normalized laser pulse profile which is approximated as a Gaussian function with a 500 ps pulse width.

*Boundary Conditions*

Surface recombination velocities (SRV) are defined at both the top surface and the back surface (i.e. glass substrate side) using the equation described elswhere[14,16] and shown below:

$$k_B T \mu_n \frac{\partial n}{\partial z}\Big|_{z=0,d} + \mu_n n\big|_{z=0,d} E_n = -\frac{np - n_0 p_0}{n S_{front,back}^{-1} + p S_{front,back}^{-1}}\Big|_{z=0,d} \qquad (S18)$$

*Summary of Drift-Diffusion Simulation Parameters*

| Parameter | Value |
|---|---|
| **Thickness** | 500 nm |
| **Absorption Coefficient** | $2\text{x}10^5$ cm$^{-1}$ ($\lambda_{exc}$ = 405 nm) [17] |
| **Excitation Intensity** | $1\text{x}10^{10}$ photons cm$^{-2}$ per pulse |
| **Laser Pulse Width** | 500 ps |
| **$k_1$ (first-order recombination rate constant)** | $3\text{x}10^4$ s$^{-1}$ |
| **$k_2^{int}$ (internal second-order rate constant)** | $5\text{x}10^{-10}$ cm$^3$ s$^{-1}$ |
| **Effective Density of States** | $3.5\text{x}10^{18}$ cm$^{-3}$ |
| **Electron and Hole Mobility** | 5 cm$^2$ V$^{-1}$s$^{-1}$ (i.e. D = 0.13 cm$^2$ s$^{-1}$) [18] |
| **Doping Concentration** | $1\text{x}10^5$ cm$^{-3}$ [17] |
| **Surface Recombination Velocity (SRV)** | *varies* cm s$^{-1}$ |
| **Temperature** | 300 K |
| **Defect Energy (midgap)** | $E_g$/2 (~0.79 eV) |

**Table S1.** Summary of drift-diffusion simulation parameters.

**Radiative Lifetime Calculations**

In the low-injection regime, the radiative lifetime of a semiconductor is limited by the background doping density, $N_A$.[19] The doping density sets a natural upper limit for the effective lifetime, which cannot exceed the radiative limit value.[20]

$$\tau_{rad} = \frac{1}{k_2^{int} N_A} \tag{S19}$$

For perovskites, $N_A$ is often cited to be $\sim 4.6 \times 10^{12}$- $4 \times 10^{13}$ cm$^{-3}$.[21-24] We use the $k_2^{int}$ value extracted from our globally fit time-resolved PL data for the control film, $k_2^{int} = 5 \times 10^{-10}$ cm$^3$ s$^{-1}$. Using these values, we approximate the radiative lifetime to be between $\sim 60$ - $450$ $\mu$s.

**Surface Recombination Velocity Calculations**

The surface recombination velocity of the perovskite active layers interfaced with the hole transport layer, Spiro-OMeTAD was determined following the procedure described by Wang *et al.* and Olson *et al.*.[19,25,26] In our experiments, the laser excitation spot size ($\sim 40$ $\mu$m, 1/e$^2$ width) is much larger than typical charge carrier diffusion lengths (i.e. $\sim$ $\mu$m's),[27] so the diffusion equation can be simplified to one dimension. Briefly, the surface recombination velocity of the top surface can be calculated analytically using the equation:

$$SRV = \frac{d}{\left(\frac{1}{\tau_{eff}} - \frac{1}{\tau_{rad}}\right)^{-1} - \frac{4}{D}\left(\frac{d}{\pi}\right)^2} \tag{S20}$$

Where $d$ is the film thickness, $\tau_{eff}$ is the measured effective decay determined from single or biexponential fitting, $\tau_{rad}$ is the bulk radiative lifetime defined as $\tau_{rad} = 1/k_2 n_d$ – here $n_d$ is the background doping density, and $D$ is the diffusion coefficient.

This assumes the bottom SRV is negligible (i.e. $\sim 0$), which we believe to be a valid assumption considering passivation of just the top surface with n-trioctylphosphine oxide (TOPO) can yield internal PLQE's as high as 97.1%.[8] The SRV values calculated using this equation are also consistent with a slightly modified version of a similar equation described in the work of Olson *et al.*.[19] We highlight that $\tau_{rad} \gg \tau_{eff}$ for most perovskite samples interfaced with a hole transport layer, and therefore the determination of this value (and the assumption of $n_d$) does not significantly impact the extracted SRV values presented in the main article.

**Radiative V$_{oc}$ and Implied V$_{oc}$ Calculations**

We calculate the radiative open circuit voltage ($V_{OC}^{rad}$) of our devices using the relation outlined by Rau[28] and the certified external quantum efficiency (EQE$_{PV}$) reported in Figure S26.

$$V_{OC}^{rad} = \frac{kT}{q} ln\left(\frac{J_{SC}}{J_{0,rad}} + 1\right) = \frac{kT}{q} ln\left(\frac{\int_{E_\gamma} EQE_{PV}(E) \phi_{AM1.5}(E) dE}{\int_{E_\gamma} EQE_{PV}(E) \phi_{BB}(E) dE} + 1\right) \tag{S21}$$

where $k$ is the Boltzmann constant, and $T$ is the temperature $q$ is the elementary charge, $J_{SC}$ is the short circuit current, J$_{0,rad}$ is the equilibrium radiative dark saturation current, $E$ is the photon energy incident on the cell's surface, $\phi_{AM1.5}$ is the Air Mass 1.5, global-tilt solar irradiance spectrum.

We obtain a $V_{OC}^{rad}$ = 1.276 V, which is similar to values calculated for other FAPbI$_3$ based devices.[29]

The quasi Fermi level splitting, which is often referred to as the implied open circuit voltage ($V_{OC}^{imp.}$), is calculated using the following equation[30]:

$$V_{OC}^{imp.} = V_{OC}^{rad} + \frac{kT}{q}ln(PLQE) \tag{S22}$$

**Explanation of PL Behaviour of 2D/3D Perovskite Heterostructures**

Surface energy potentials induced by chemical treatment and applied biases have been shown to significantly impact the rate of radiative recombination of InP,[31-33] CdS,[34] and GaAs.[35] The impact of surface treatments on the rate of radiative recombination is relatively unexplored for metal halide perovskites, especially in the case of organic salts (i.e. HABr or PEAI) and the formation of 2D/3D heterostructures. This understanding is critical to distribution of charge carriers through the material stack and therefore the operation of nearly all perovskite-based optoelectronic devices. Indeed, just from performing simple time-resolved PL decay measurements and exciting the sample from the front and back side at different laser intensities, it is clear that the 2D/3D heterostructure leads to unique charge carrier dynamics distinct from bare 3D films (see Figures S20 and S22). Here we summarize the distinct behaviours and perform a set of physically-informed numerical simulations that, for the first time, reproduce these observations:

**1)** Excitation from the front surface (i.e. 2D side) shows a very rapid decrease in the PL intensity – this drop can be as large as 2 orders of magnitude (see Figures S20 and S22). This drop in intensity is dependent on the concentration of the surface treatment (see Figure 1c).

**2)** Excitation from the back side (i.e. 3D side) results in less of a rapid decay and instead the PL decay can be significantly extended, relative to the control sample (see Figure 1). The extension in PL lifetime is dependent on the concentration of the surface treatment (see Figure S1).

**3)** Excitation from the back side and global fitting of intensity dependent TRPL decay traces yields an effective internal radiative recombination, $k_{2,eff}^{int}$, that is much lower than what is typically observed for bare 3D films. This reduction in $k_{2,eff}^{int}$ is observed for several different 2D/3D systems including HABr and PEAI treated films (see Figures S1, S4, and S20).

*Numerical Model Description*

In order to better understand **3)**, we perform a 1-D numerical simulation of the electron and hole densities as a function of time and film depth. Figure S21 shows the energy level diagrams used for the simulation. In the control case, flat bands are used, and in the treated case 100 meV and 200 meV band bending values are used. The energy grading values are defined by difference in energy between flat band region and maximum energy of the band (i.e. difference in energy value at a depth of 0 nm minus the value at a depth of 500 nm). For these specific simulations, we seek to isolate how band bending and the spatial separation of electron and holes impact the PL emission rate. Therefore, we used all the parameter values described in Table S1, but set the surface recombination velocities for the front and back surfaces to 0 in order to remove the impact of this term on the local electron and hole densities. In addition, the electron mobility, $\mu_e$, was set to 0.23 cm$^2$ V$^{-1}$ s$^{-1}$ and the hole mobility was set to 17.5 cm$^2$ V$^{-1}$ s$^{-1}$ which are measured values reported by Zhai *et al.* for a mixed formamidinium(FA)/methylammonium(MA) perovskite system.[36] We perform this simulation using the same parameters for Figure S21b, c, and d and only change the shape of the energy level diagram – to isolate the impact of band bending. Figure S21 shows that the PL intensity decreases across the film thickness with the energy gradings and especially at the top surface where electrons are being repelled and holes accumulate.

*Impact of Band Bending on Effective Bimolecular Recombination Rate*

In order to quantify the change in overall PL emission rate, we take the average PL across the whole PL intensity map. For the flat band scenario, the average PL (PL$_{avg.}$) is 1.3x10$^{23}$ photons cm$^{-3}$ s$^{-1}$, for 100 meV of band bending PL$_{avg.}$ = 1.0x10$^{23}$ photons cm$^{-3}$ s$^{-1}$, and for 200 meV, PL$_{avg.}$ = 2.2x10$^{22}$ photons cm$^{-3}$ s$^{-1}$. As $k_2^{int}$ is an intrinsic material parameter and was kept constant for these simulations, the origin of PL$_{avg.}$ changing for the band bending scenarios is due to the local product of the electron and hole densities decreasing (i.e. $PL \sim k_2^{int}np$). Although these changes PL emission rate vary locally (see Figure S21c and d), we use the average values to quantify the extent of electron-hole separation. These values come out as a constant, and are absorbed into an effective recombination rate constant, $k_{2,eff}^{int}$. Therefore, we use the PL$_{avg.}$ values and their relative ratios to quantify the impact on $k_{2,eff}^{int}$. We note that the reference value is 5x10$^{-10}$ cm$^3$s$^{-1}$ for the flat band scenario, which we extracted from global fits to the intensity-dependent TRPL of the control sample (see Figure S1). Importantly, for the 200 meV band bending scenario, we observe a significant reduction in the $k_{2,eff}^{int}$ value, which matches well with $k_{2,eff}^{int}$ extracted from the global fits the HABr and PEAI treated samples shown in Figures S4 and S20. We highlight that the change in PL$_{avg.}$, and hence $k_{2,eff}^{int}$, is highly dependent on the magnitude and shape of band bending as well as other factors such as the electron and hole mobilities and film thickness. These simulations therefore capture the important photophysical processes, and could be further improved with accurate measurements of the band structure as well as electron and hole mobilities.

*Simulation of TRPL Kinetics when Excited from Front and Back Surfaces*

Next, we test the numerical model's ability to capture observations **1)** and **2)**. For comparison, we measured the TRPL dynamics when photoexciting ($\lambda_{exc}$ = 405 nm, 62.5 kHz, 20 nJ cm$^{-2}$ per pulse, $N_0 \sim$ 6x10$^{14}$ cm$^{-3}$) the front and back surfaces of a control, 10 mM HABr, and 50 mM HABr treated samples (Figure S23a and b). We perform the same simulations for the flat band and band bending scenarios shown in Figure S21, but incorporate a front surface recombination velocity (SRV) of 25 cm s$^{-1}$ and keep the back SRV = 0 cm s$^{-1}$, consistent with the top surface being the primary source of non-radiative defects.[8,37] Figure S23c, shows the presence of a fast drop in PL that increases in magnitude with larger band bending and when the sample is excited from the front surface. Contrarily, when the sample is excited from the back side, Figure S23d shows that the PL lifetime becomes longer with increasing band bending. Both of these observations in the simulated data are consistent with experimental measurements in Figure S23a and b and are a result of field-induced spatial separation of electrons and holes.

We highlight that the magnitude of the fast PL decay feature when the sample is excited from the front surface and the determination of $k_{2,eff}^{int}$ when photoexciting the sample from the back side can be used as quantitative metrics to describe the extent of band bending and electron and hole separation. For example, a larger reduction in the PL when excited from the front side and a lower $k_{2,eff}^{int}$ translate to larger band bending, according to our numerical model predictions. This conclusion is consistent with higher concentrations of surface treatments leading to larger band bending and slower radiative recombination (see Figures S21 and S23).

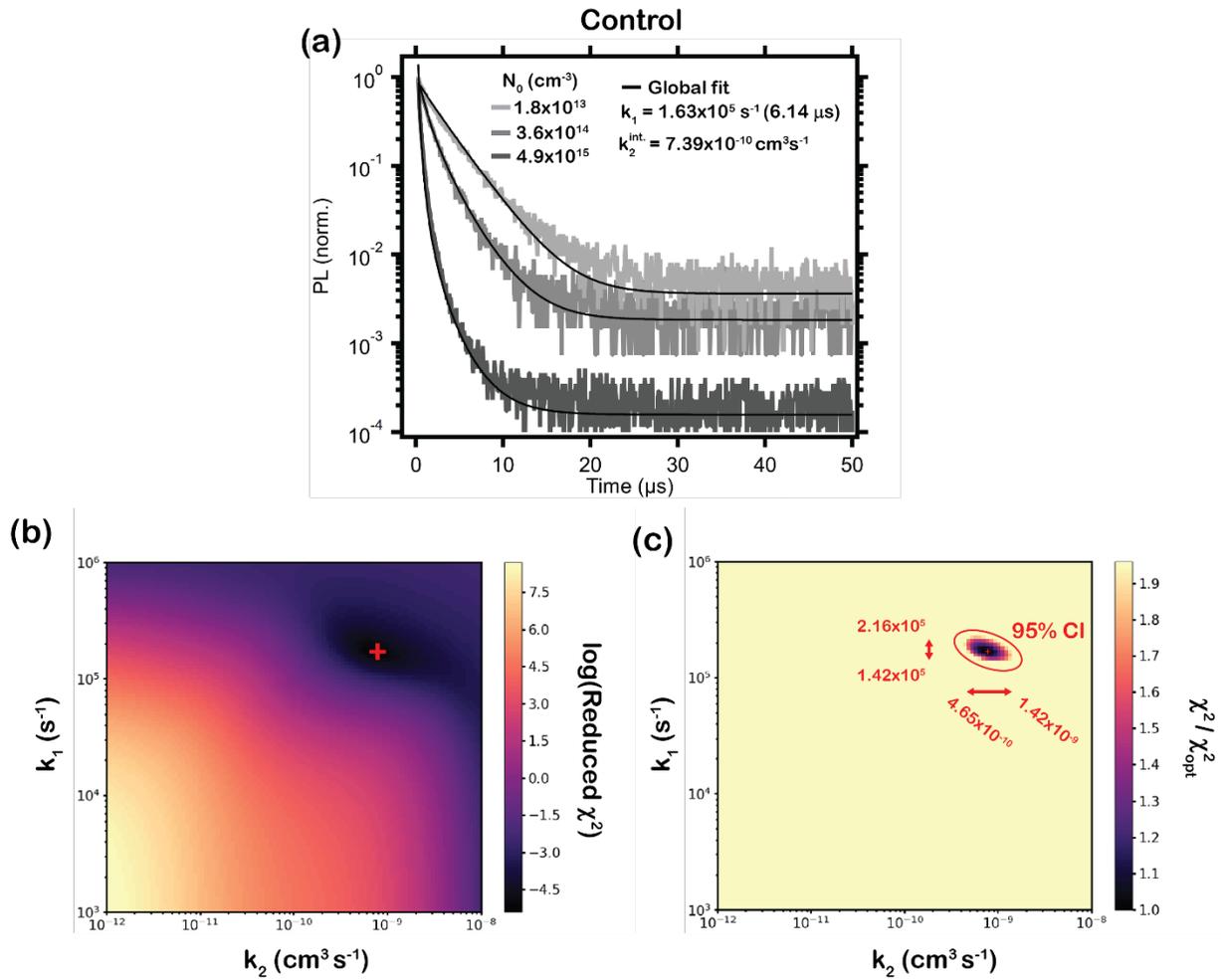

**Figure S1.** a) Time-resolved photoluminescence decay traces of a control sample measured at different excitation densities, along with global fits and extracted first-order, $k_1$, and second-order, $k_2$, recombination rate constants. b) Reduced $\chi^2$ surface of the free variables $k_1$ and $k_2$, where the red plus sign marks the minimum of this surface and the optimized rate constants. c) Support plane analysis of the $k_1$ and $k_2$ parameters, where values are thresholded at a confidence interval of 95%. The arrows beside and below the circled region show the upper and lower limits of a 95% CI for each fitted parameter.

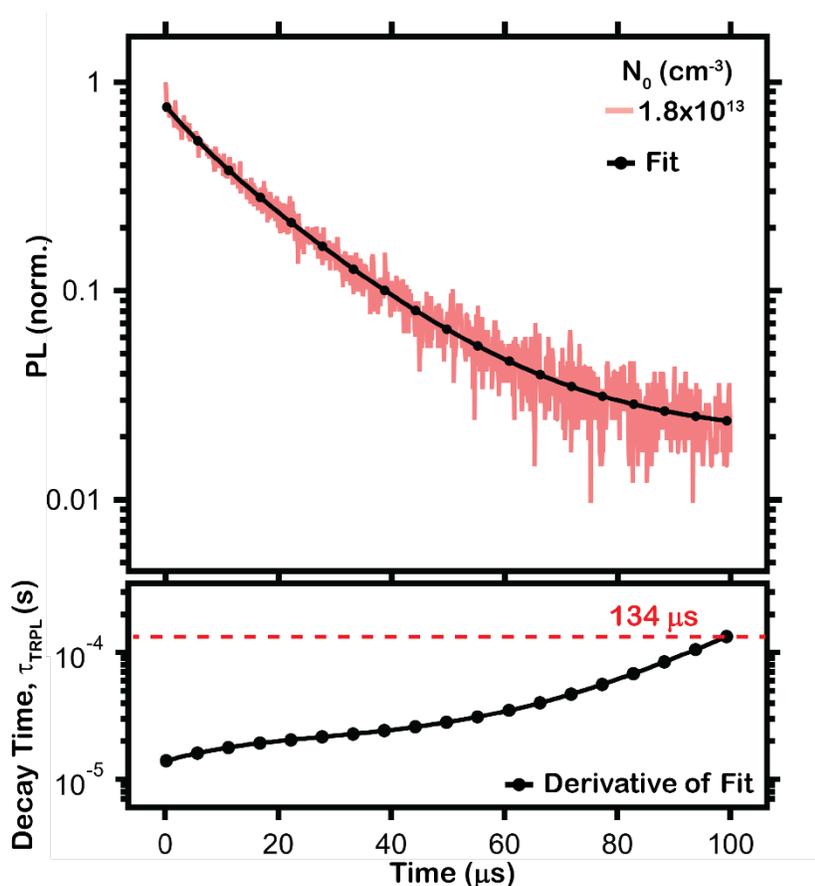

**Figure S2.** Upper panel. Time-resolved photoluminescence decay trace of a 50 mM HABr treated perovskite sample measured at a low excitation fluence with an initial carrier density ($N_0$) of $1.8 \times 10^{13}$ cm$^{-3}$. The solid black line is a multiexponential fit used to calculate the derivative of the PL decay trace. Lower panel. Differential decay time determined by taking the derivative of the fit in the upper panel and using the equation $\tau_{TRPL} = \left(-\frac{dln(PL_{int.})}{dt}\right)$.[38] We note that we do not differentiate between being in the low or high-level injection regime as observing truly monoexponential behavior from bulk TRPL measurements is unlikely due to the kinetics being composed of an ensemble of decay rates.[39,40]

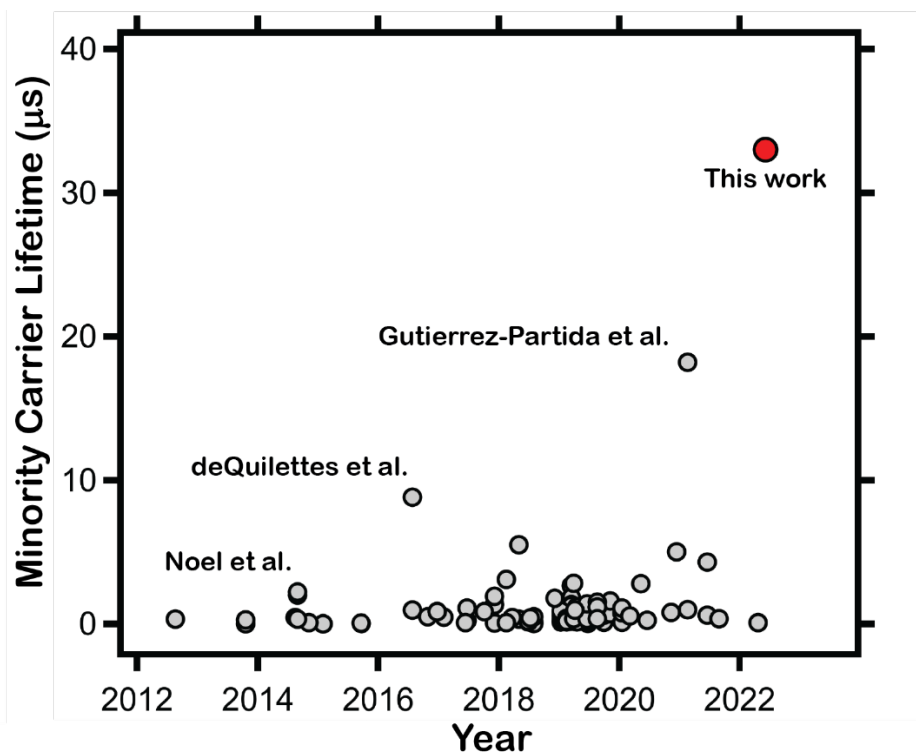

**Figure S3.** Literature survey of the reported minority carrier lifetimes of perovskites over the last decade. The average lifetime value is $1.06 \pm 2.13$ $\mu$s (number of samples, N = 102). This work demonstrates over an order of magnitude improvement in the average value. See Supplementary Table S2 for references.

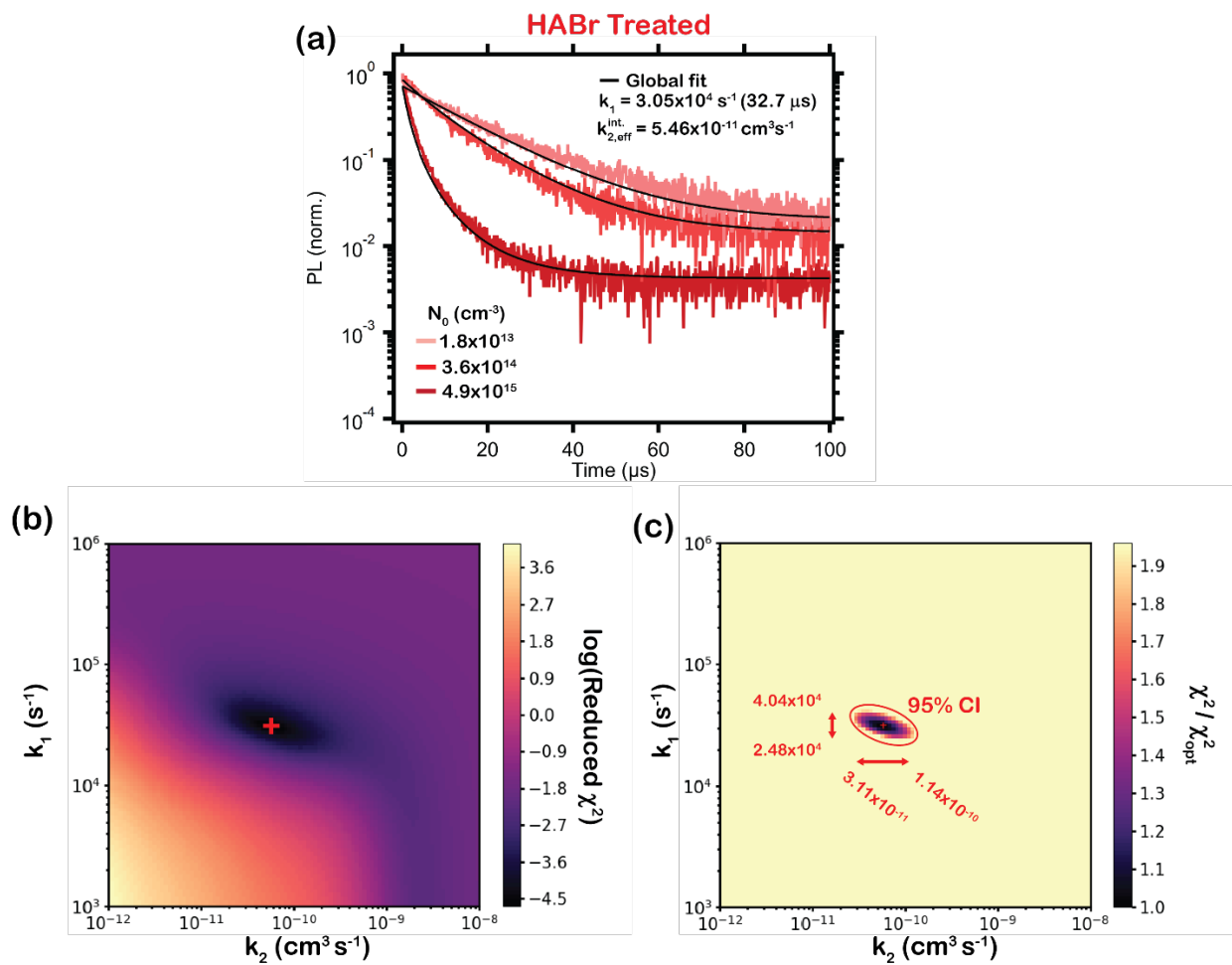

**Figure S4.** a) Time-resolved photoluminescence decay traces of a 50 mM HABr treated sample measured at different excitation densities, along with global fits and extracted first-order, $k_1$, and second-order, $k_2$, recombination rate constants. b) Reduced $\chi^2$ surface of the free variables $k_1$ and $k_2$, where the red plus sign marks the minimum of this surface and the optimized rate constants. c) Support plane analysis of the $k_1$ and $k_2$ parameters, where values are thresholded at a confidence interval of 95%. The arrows beside and below the regions show the upper and lower limits of a 95% CI for each fitted parameter.

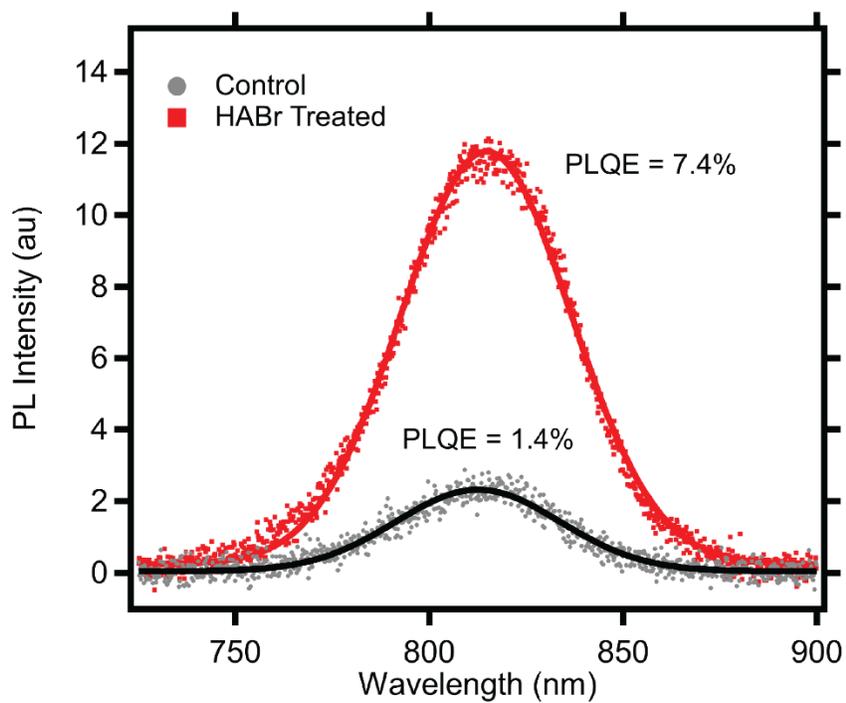

**Figure S5.** Photoluminescence (PL) and calculated PL quantum efficiency (PLQE) for control and HABr treated films measured at 1-sun equivalent absorbed photon flux using a 532 nm continuous wave laser.

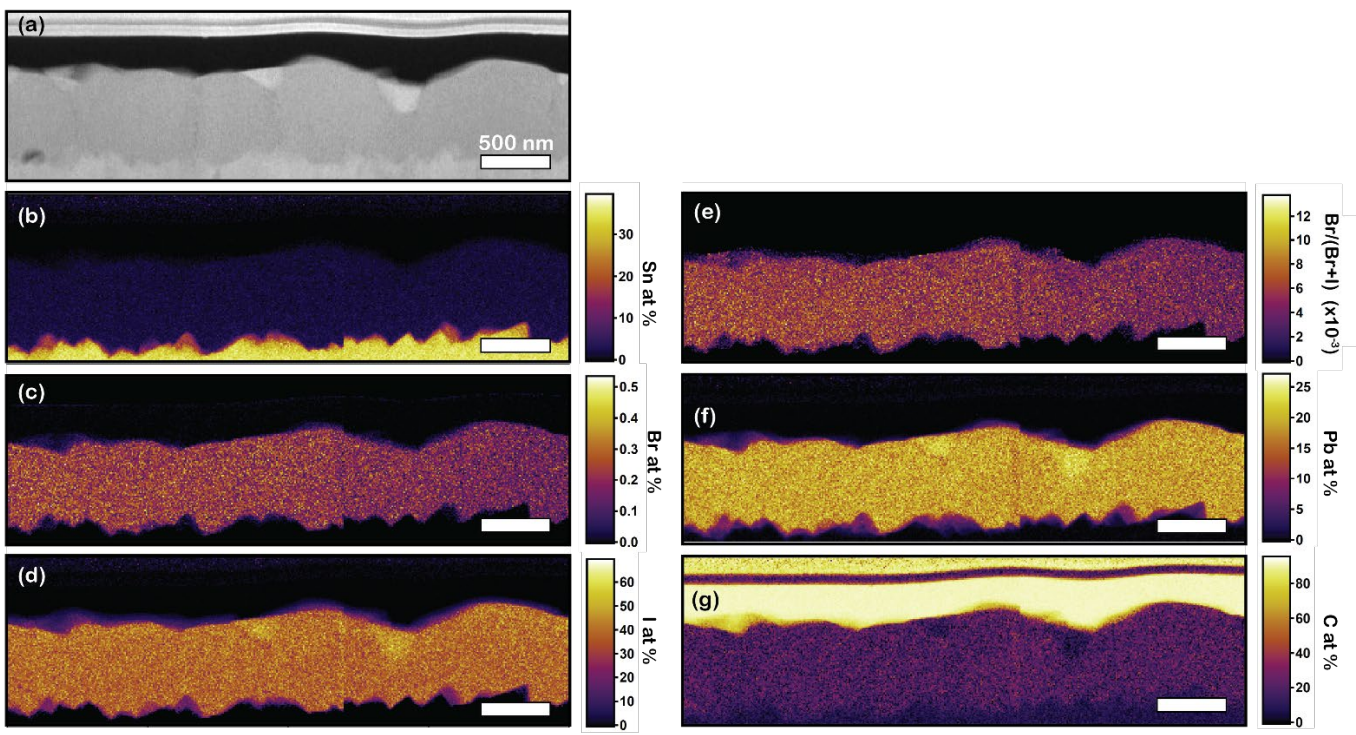

**Figure S6.** a) Cross-sectional STEM-HAADF image of a control sample. Correlated quantitative STEM-EDX maps of b) Sn, c) Br, d) I, e) Br/(Br+I), f) Pb, and g) C. Pseudo-color scale bars for b-g) are shown to the right of each image.

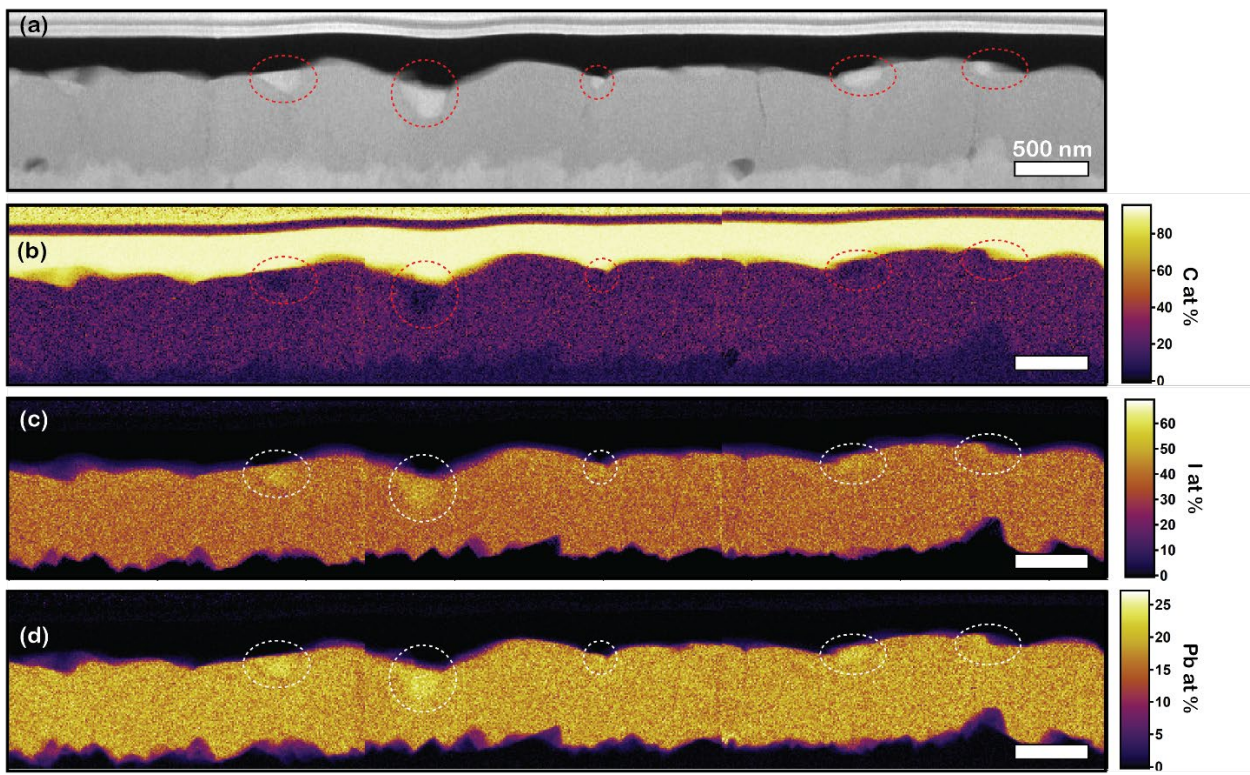

**Figure S7.** a) Cross-sectional STEM-HAADF image of a control sample. Correlated quantitative STEM-EDX maps of b) C, c) I, and d) Pb. Circled regions in a) show higher intensity, indicative of a higher effective atomic number, which is correlated to lower C content and higher I and Pb. These regions are concentrated at grain boundaries and are likely $PbI_2$. Pseudo-color scale bars for b-d) are shown to the right of each image.

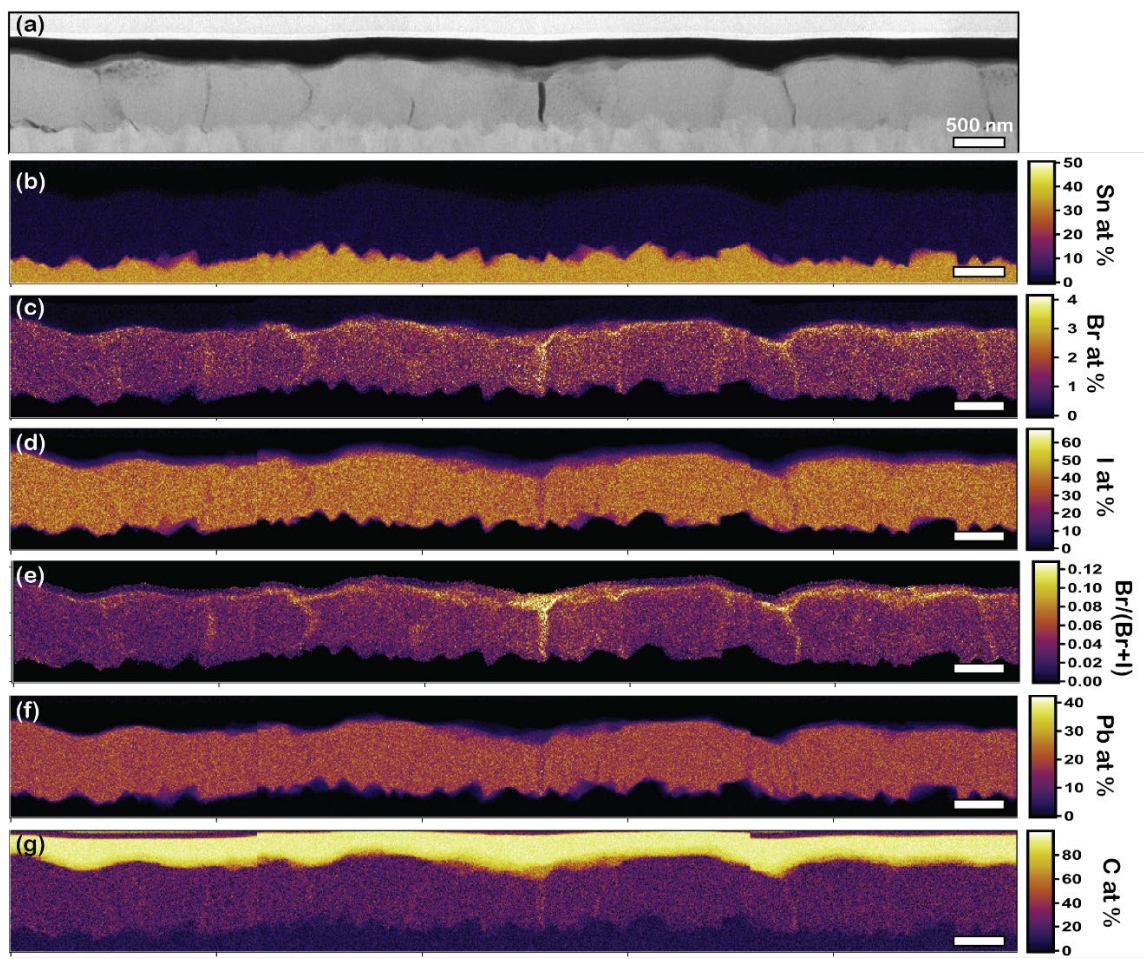

**Figure S8.** a) Cross-sectional STEM-HAADF image of a 50 mM HABr treated sample. Correlated quantitative STEM-EDX maps of b) Sn, c) Br, d) I, e) Br/(Br+I), f) Pb, and g) C. Br is concentrated at the top surface as well as grain boundaries. Pseudo-color scale bars for b-g) are shown to the right of each image.

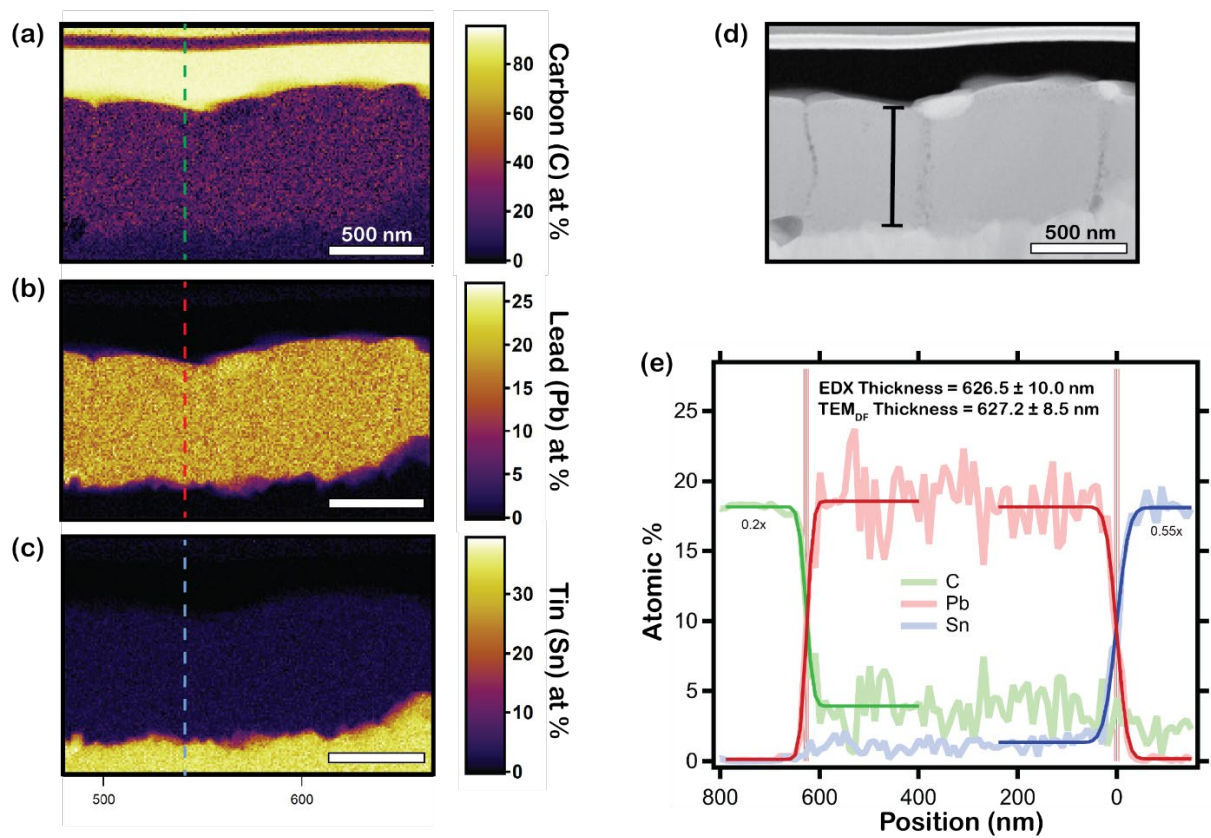

**Figure S9.** Cross-sectional STEM-EDX maps of a) C, b) Pb, and c) Sn of a control sample, along with dashed lines to identify where line profiles were taken to determine the interface positions and perovskite layer thickness. d) corresponding cross-sectional STEM-HAADF image of the same region and line, showing where the thickness was measured using the spatially calibrated image. e) atomic % line profiles along with fits to each trace using an error function. The thicknesses determined using the STEM-HAADF image and the STEM-EDX line profile analysis were within error of each other, implying that this method can be used to accurately determine the interface positions. Pseudo-color scale bars for a-c) are shown to the right of each image.

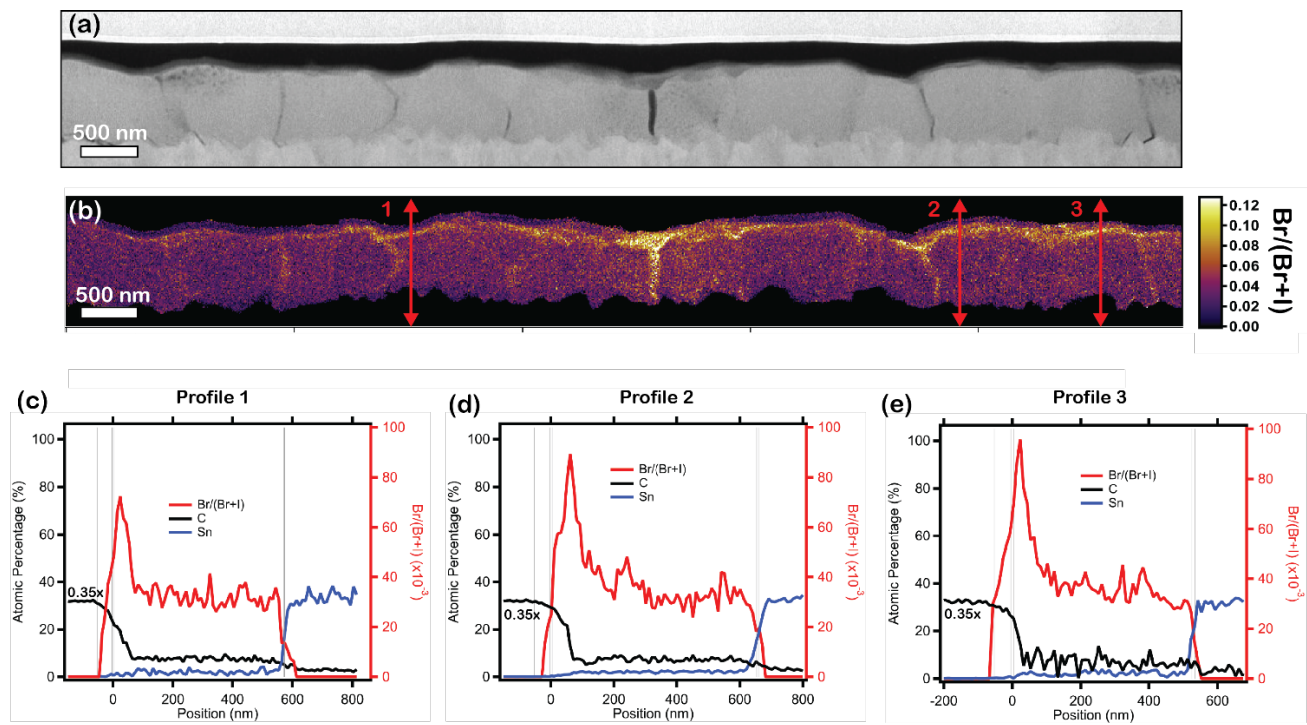

**Figure S10.** a) Cross-sectional STEM-HAADF image of a 50 mM HABr treated sample. b) Correlated quantitative STEM-EDX maps of the Br/(Br+I) ratio. c-e) line profiles of three different regions across the sample showing the Br peak is in the 3D bulk layer and not in the 2D layer. Vertical lines in c), d), and e) are the interface positions determined using Equation S1, along with the measured film thicknesses. Pseudo-color scale bar for b) is shown to the right of the image.

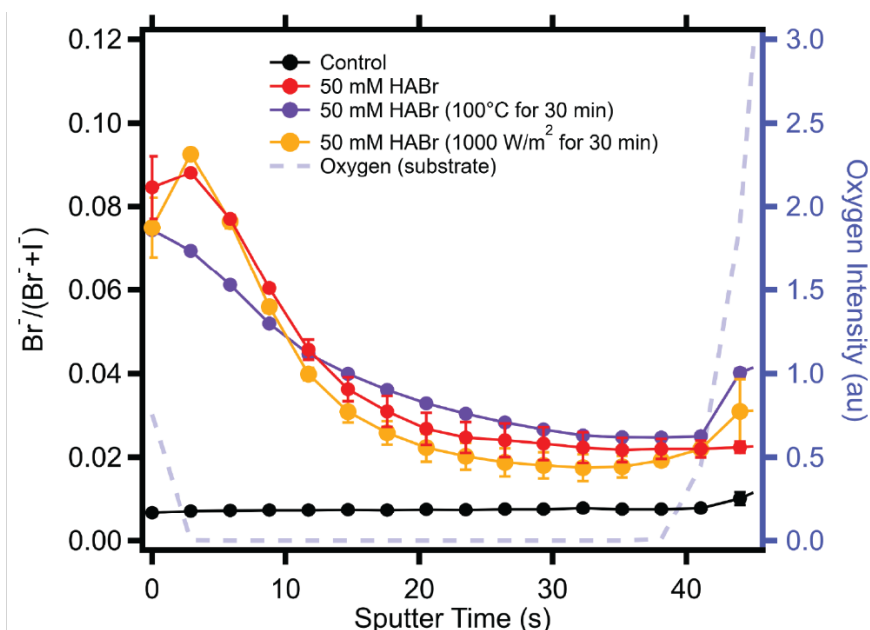

**Figure S11.** Depth-dependent time of flight secondary ion mass spectrometry (ToF-SIMS) showing the Br⁻ / (Br⁻ + I⁻) ratio of a control, a 50 mM hexylammonium bromide (HABr) treated sample, and 50 mM HABr sister samples exposed to 30 min of light (1000 W/m²) and heat (100 °C). The ToF-SIMS oxygen (O⁻) intensity (dotted blue trace) is shown on the right axis in to signify the interface with the glass (i.e. $SiO_x$) substrate.

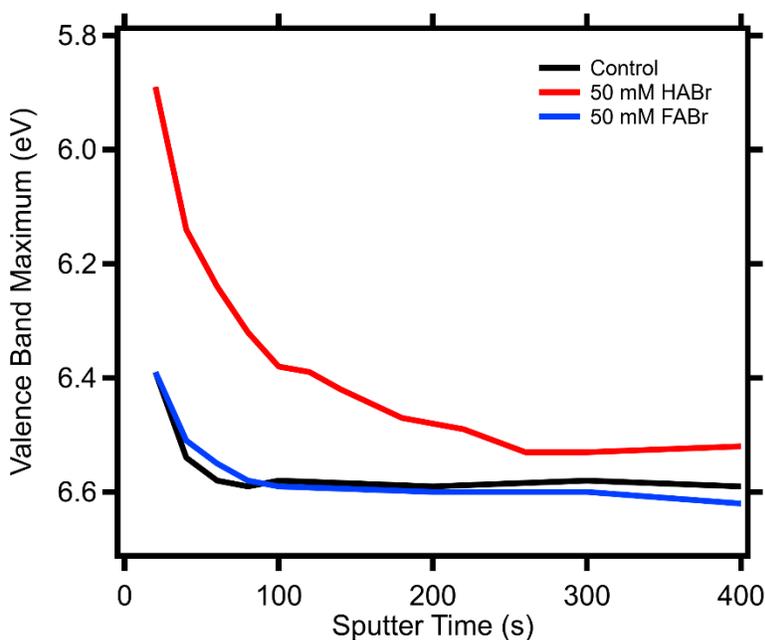

**Figure S12.** Valence band maximum determined from depth-dependent ultraviolet photoemission spectroscopy (UPS) of a control, 50 mM HABr treated sample, and 50 mM formamidinium bromide (FABr) treated sample. The control and 50 mM FABr treated sample appear very similar.

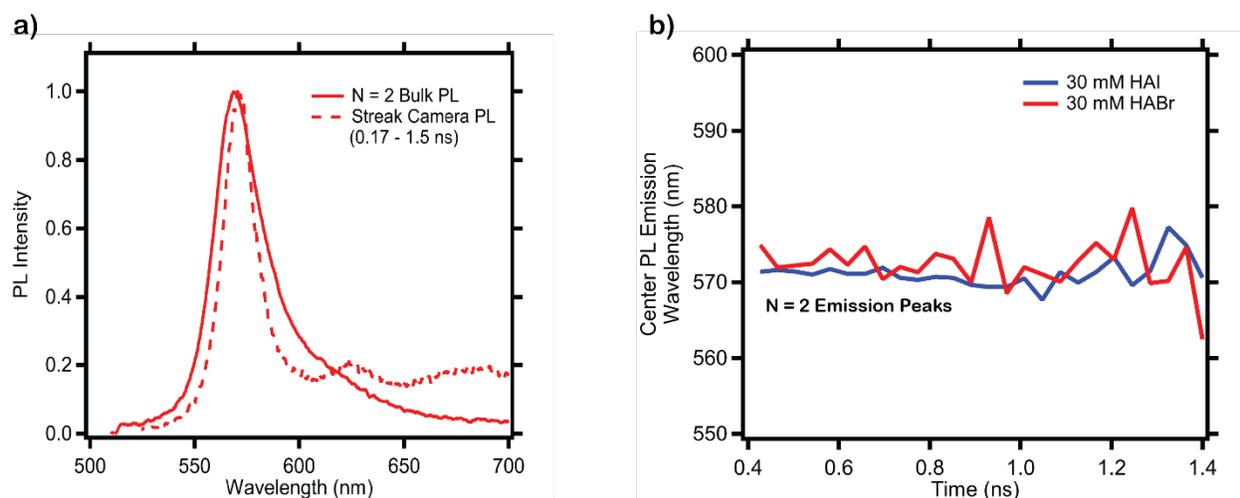

**Figure S13.** a) Photoluminescence spectra of a bare $HA_2FAPb_2I_7$ (n = 2) film on glass (solid red line) compared to the PL emission profile of the high energy peak of a HABr treated film. Data was collected using a streak camera and is the same data set in Figure 3b integrated over 0.17-1.5 ns. b) Center emission line profiles as a function of time for a 2D layer formed on top of a 3D perovskite using hexylammonium iodide (HAI) versus HABr. The emission peaks overlap in both scenarios, suggesting that the 2D layer formed on top of the 3D layer has negligible Br content and forms the n = 2 $HA_2FAPb_2I_7$ low dimensional perovskite even when HABr is used.

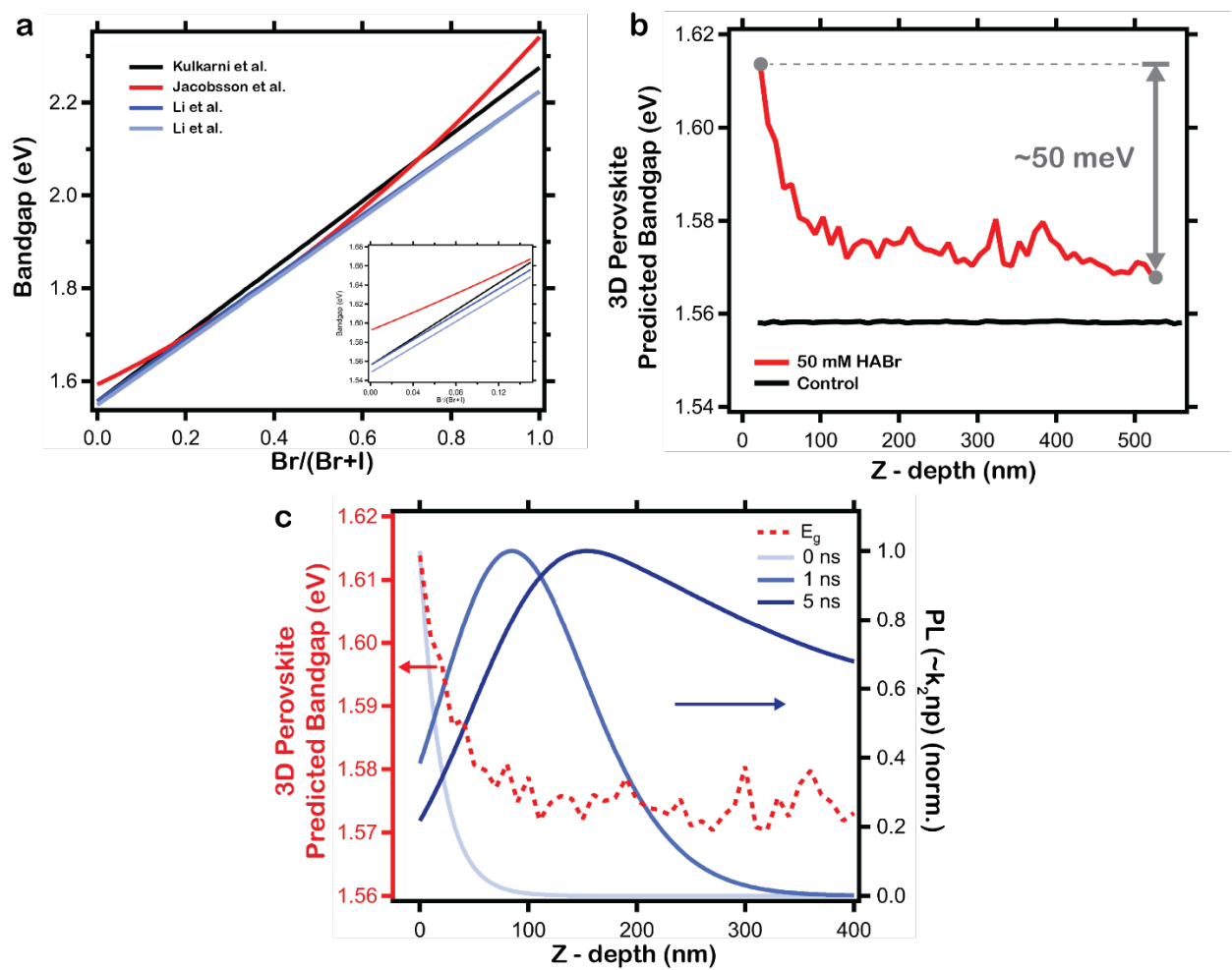

**Figure S14.** a) Perovskite optical bandgap energy as a function of Br/(Br+I) ratio based on previous reports in the literature.[41-43] Inset – zoomed in version of same data set over a smaller range of Br/(Br+I) ratios. Kulkarni *et al.* only considered methylammonium (MA)-based perovskites, therefore, the starting value was modified to match the initial values from Li *et al.*, which take into account contributions from formamidinium (FA) additions. b) Predicted bandgap of the 3D perovskite material using the quantitative Br ratio obtained from the line profiles in Figure 2e, and f in the main article. We note that an additional 70 meV offset was applied to all traces in a) (i.e. offset in the absolute y-value) in order to match the $E_g$ determined from EQE spectrum[44] (see Figure S26) with the curves from literature at a fixed BBI value. The slopes (i.e. changes in $E_g$ as a function of BBI ratio) of each trace were not affected and therefore have a negligible impact on the overall magnitude of the $E_g$ grading reported in b). c) Simulated photoluminescence profile through the 3D film thickness after 405 nm excitation from the top surface and accounting for drift from the bandgap grading (red dotted line) from b). The PL trace is a product of the instantaneous electron and hole densities, which probes different regions of the compositional gradient (i.e. bandgap) as a function of time.

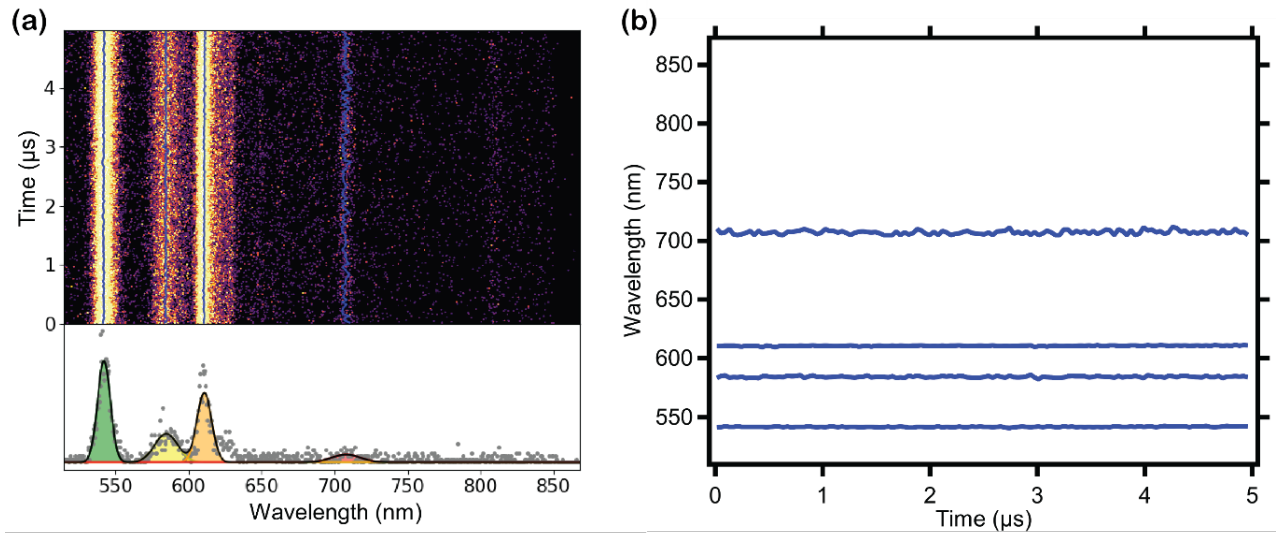

**Figure S15.** a) Streak camera image (using pseudo-color scale) of compact fluorescence lights (CFLs) with characteristic emission lines of mercury and the fluorescent phosphor coating. Spectra at each time slice were fit as a summation of Gaussian functions, where the blue line shows the center emission energy. b) Center emission as a function of time from a). We observe no apparent shifts in wavelength as a function of time, suggesting no non-linearity in the streak tube or camera tilt and therefore the shifts observed in the main text are real.

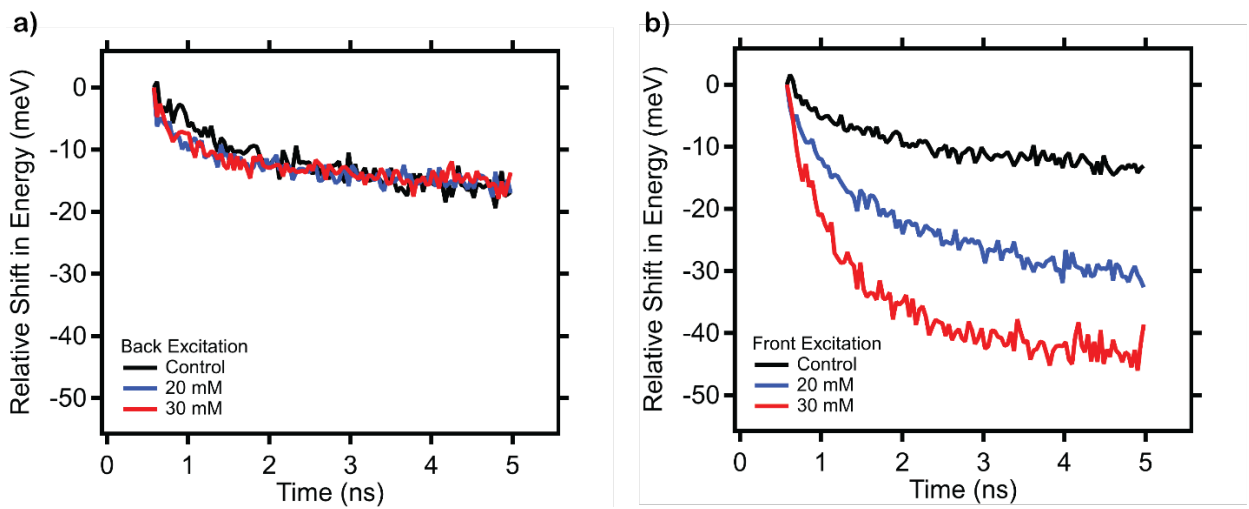

**Figure S16.** a) Shifts in PL emission energy of control, 20 mM, and 30 mM treated samples when excited from the glass side (opposite of where the 2D layer is), all showing similar shifts as a function of time. b) Shift in PL emission energy as a function of time for a control (black trace), 20 mM HABr treated (blue trace), and 30 mM HABr treated (red trace). Treatments using higher concentrations lead to larger shifts in emission energy.

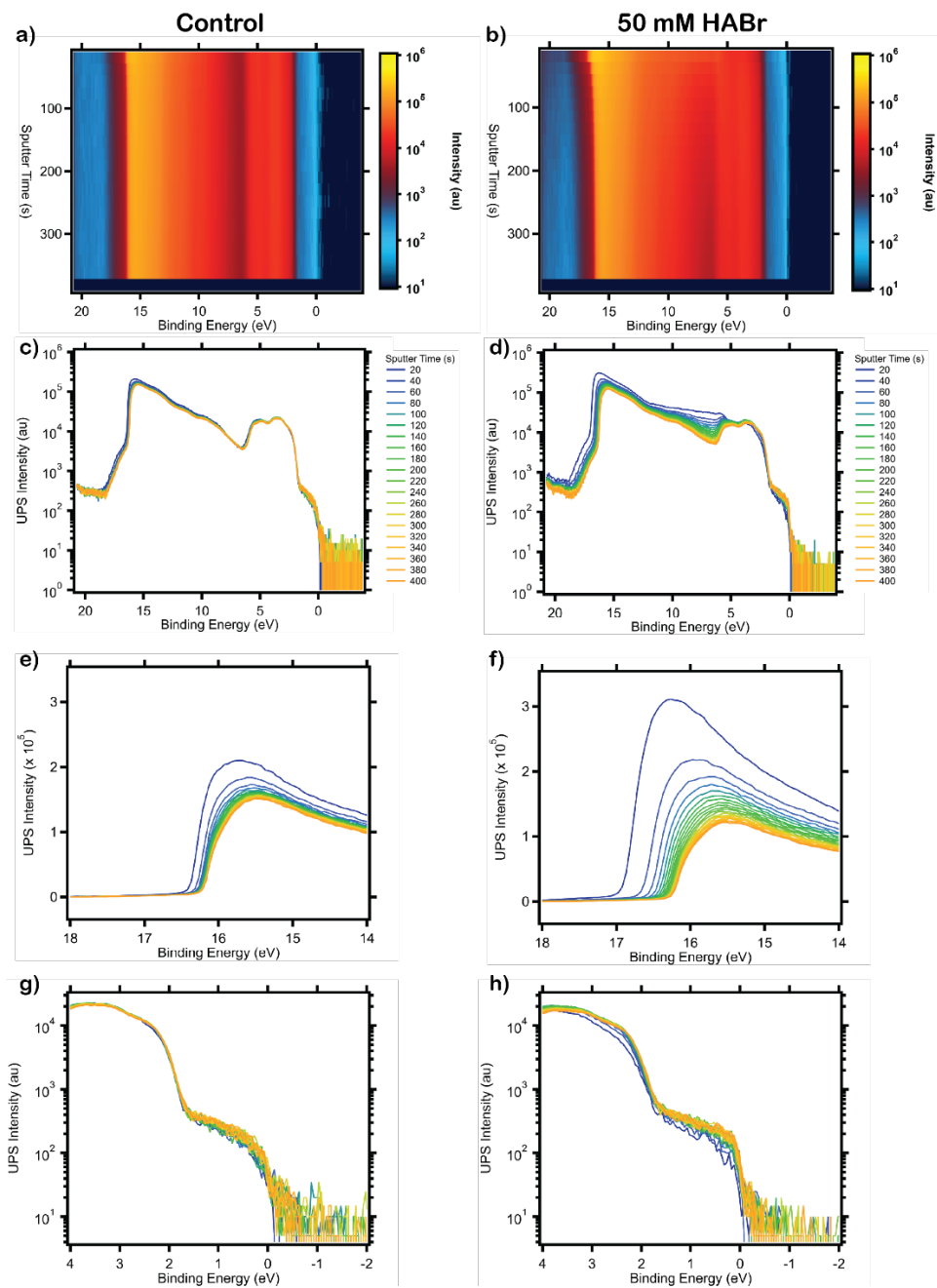

**Figure S17.** Depth-dependent ultraviolet photoemission spectroscopy (UPS) heat map of a a) control and b) 50 mM HABr treated sample. Spectral slices of the c) control and d) HABr treated samples at sputtering time intervals of 20 s. Zoomed in of c) and d) at the secondary cut off energy for the e) control and f) HABr treated sample. Zoomed in of c) and d) at the valence band minimum for the g) control and h) HABr treated sample.

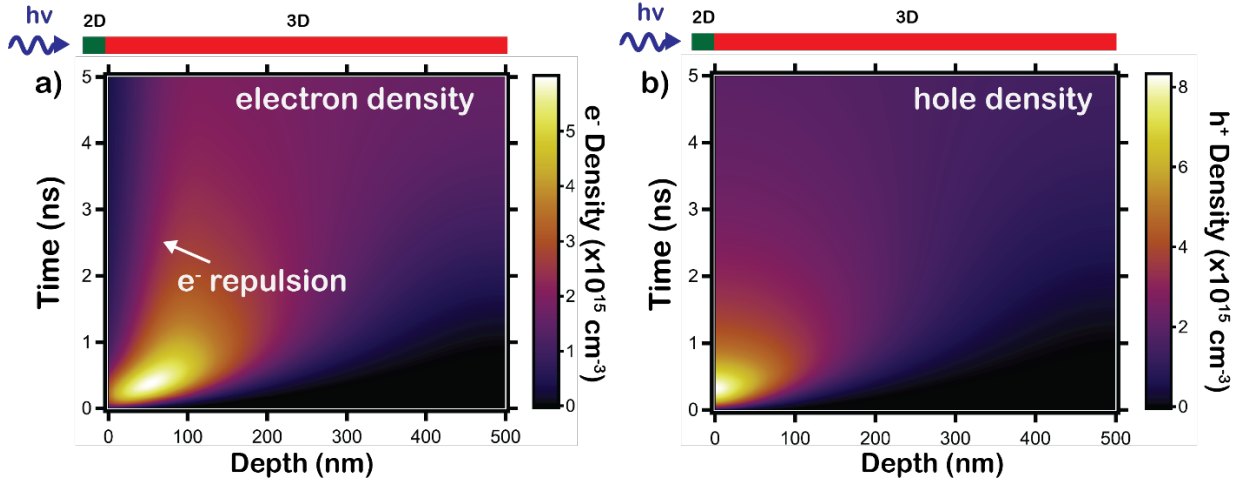

**Figure S18.** Simulated a) electron and b) hole densities as a function of time through the film thickness with a 50 meV energy grading when excited from the front (i.e. top surface/2D side). The bandgap grading in this case, which is expected to mostly impact the conduction band energy,[15] only affects the electron density at the top surface/interface and not the hole density. Simulation parameters shown in Table S1 were used, with an SRV = 25 cm s$^{-1}$ and electron and hole mobility of 5 cm$^2$ V$^{-1}$ s$^{-1}$.

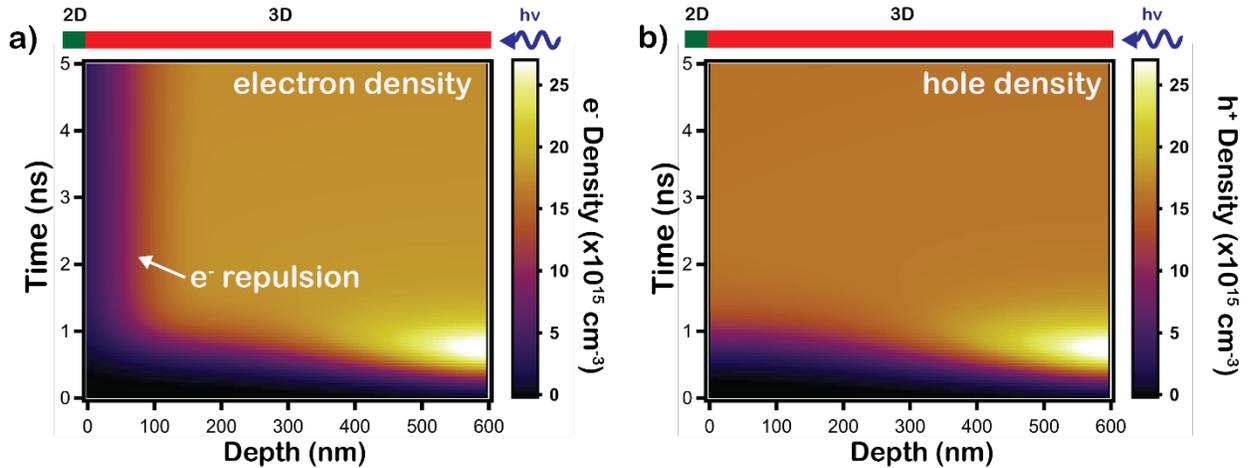

**Figure S19.** Simulated a) electron and b) hole densities as a function of time through the film thickness with a 50 meV energy grading when excited from the back (i.e. bottom surface/3D side). The bandgap grading in this case, which is expected to mostly impact the conduction band energy,[15] only affects the electron density at the top surface/interface and not the hole density. Simulation parameters shown in Table S1 were used, with an SRV = 0 cm s$^{-1}$ and electron and hole mobility of 25 cm$^2$ V$^{-1}$ s$^{-1}$, in order to emphasize the difference in electron and hole populations at the front surface over the shorter time window.

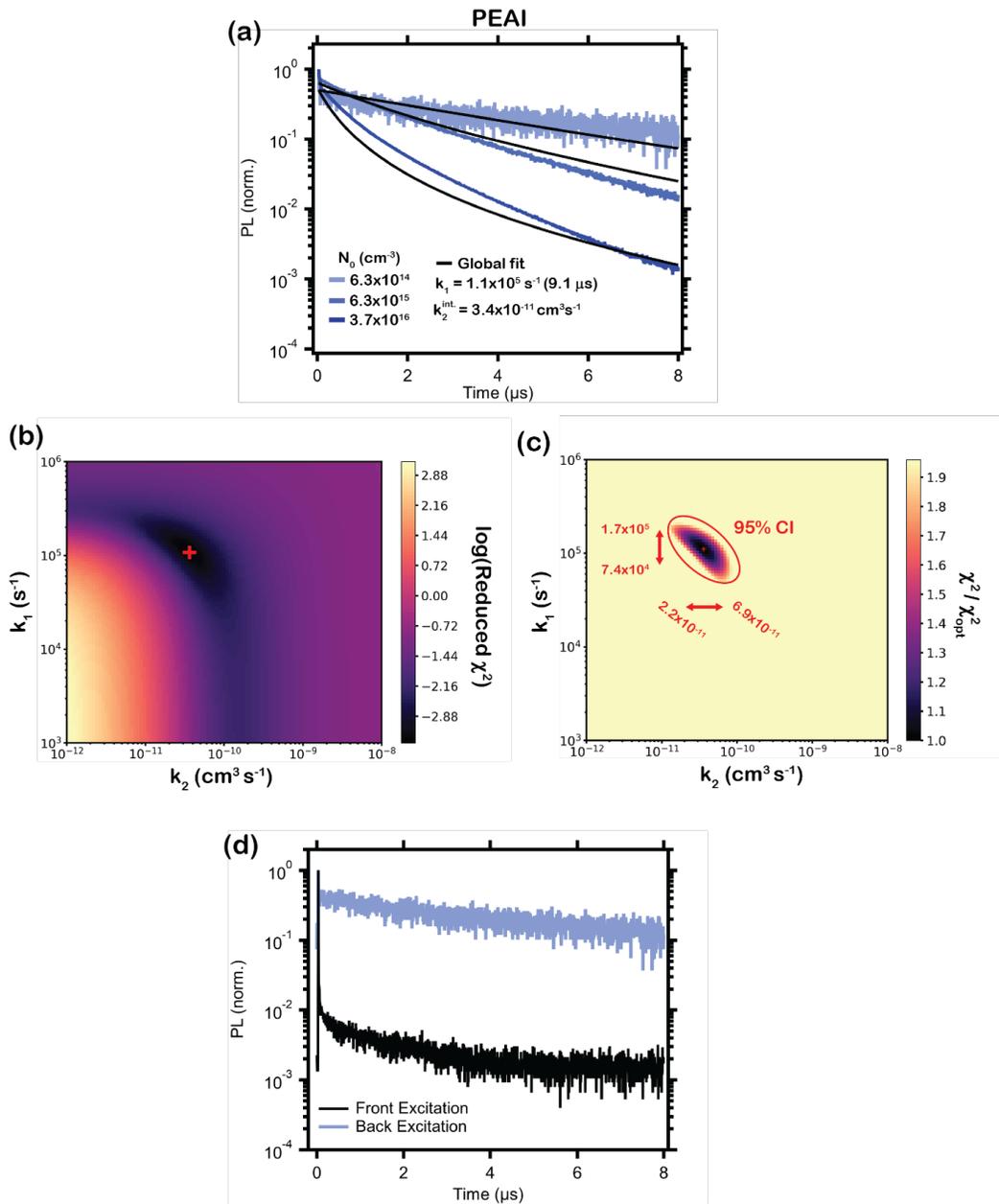

**Figure S20.** a) Time-resolved photoluminescence decay traces of a 50 mM PEAI treated sample measured at different excitation densities, along with global fits and extracted first-order, $k_1$, and second-order, $k_2$, recombination rate constants. b) Reduced $\chi^2$ surface of the free variables $k_1$ and $k_2$, where the red plus sign marks the minimum of this surface and the optimized rate constants. c) Support plane analysis of the $k_1$ and $k_2$ parameters, where values are thresholded at a confidence interval of 95%. The arrows beside and below the circled region show the upper and lower limits of a 95% CI for each fitted parameter. d) Time-resolved PL of the same PEAI sample photoexcited from the front (top surface, black) and back (glass side, light blue) at the lowest excitation fluence.

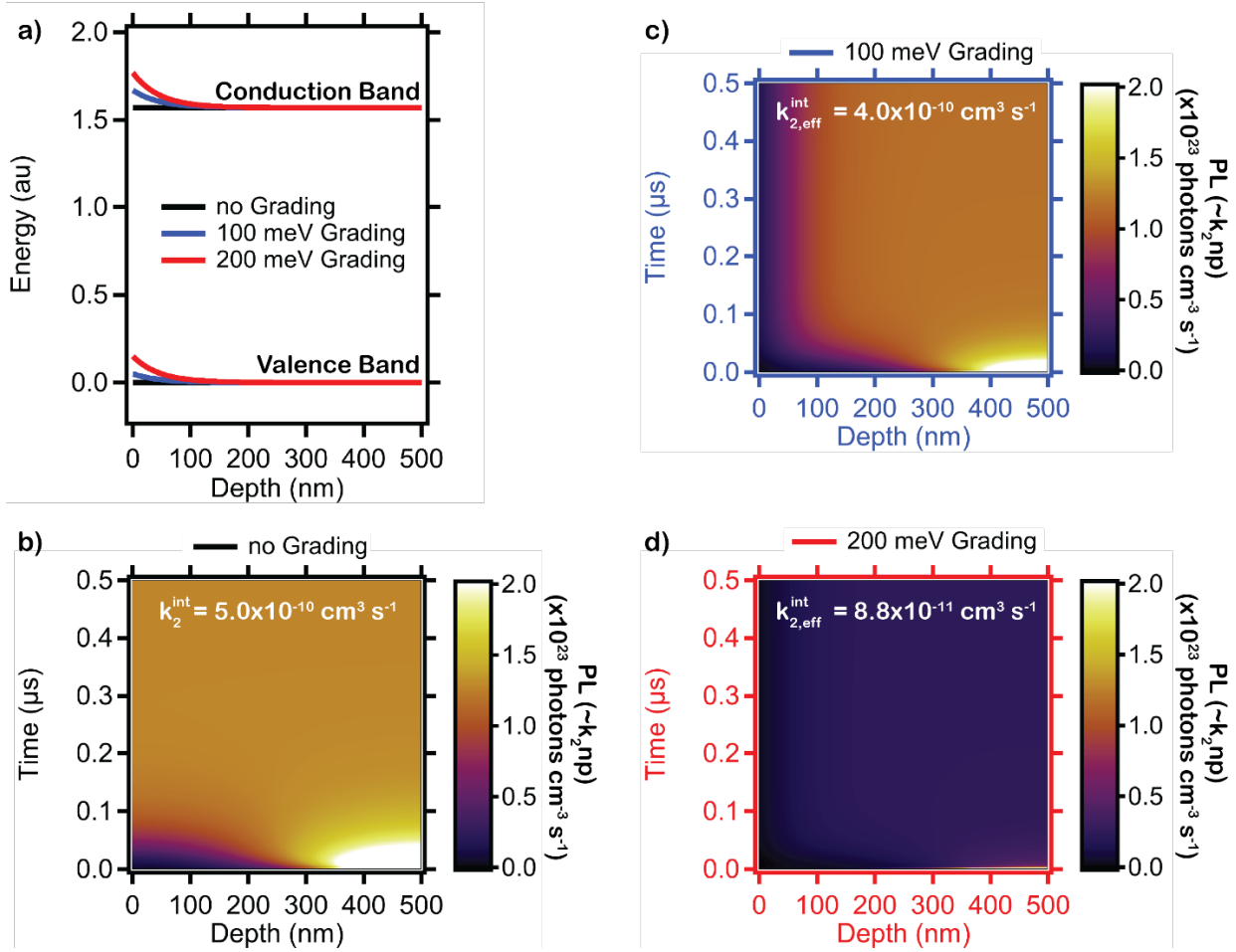

**Figure S21.** a) Simulated energy levels for valence and conduction bands with no band bending, 100 meV, and 200 meV energy gradings. Energy gradings are defined by difference in energy between flat band region (i.e. maximum value at a depth of 0 nm minus the value at a depth of 500 nm). b) Simulated photoluminescence (PL) map for the scenario with b) no band bending, c) 100 meV energy grading, and d) 200 meV energy grading. All color scale bars in b-d) are the same allowing for direct comparison of the PL emission rate for the case with no spatial separation of electron and holes pairs (b) versus the scenarios where electron and holes spatially separate (c and d). The $k_{2,eff}^{int}$ values in c) and d) capture the decrease in the radiative emission rate compared to the baseline value of $5\times10^{-10}$ cm$^3$s$^{-1}$ (see section with header "Explanation of PL Behaviour of 2D/3D Perovskite Heterostructures" for further details and simulation parameters).

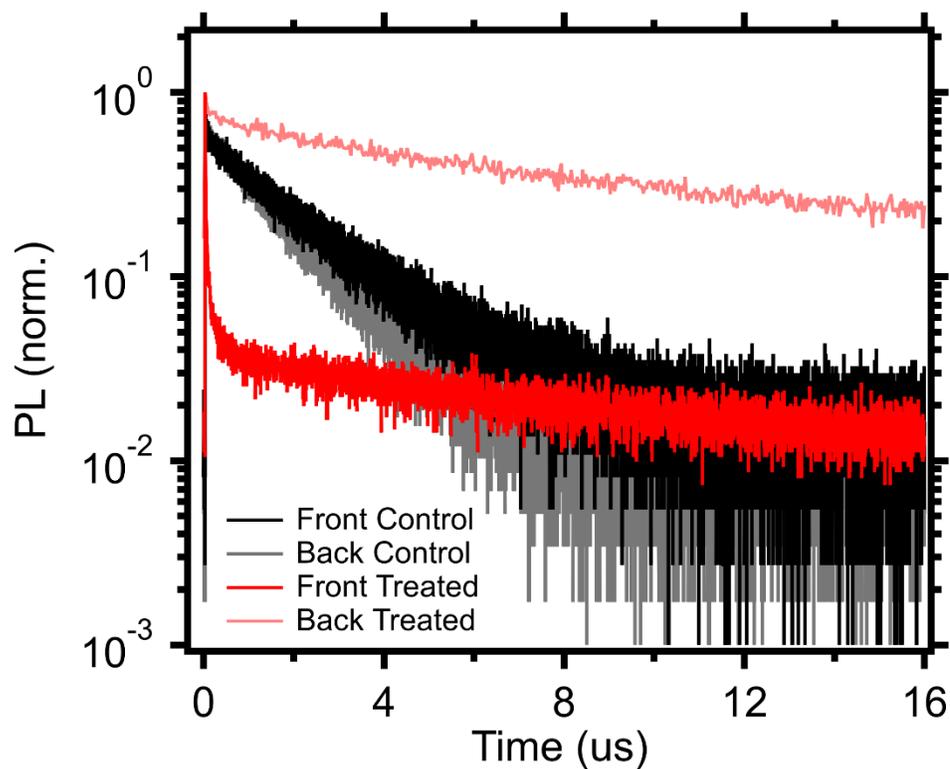

**Figure S22.** Time-resolved PL of control (black) and 50 mM HABr treated (red) samples photoexcited ($\lambda_{exc}$ = 405 nm, 62.5 kHz, 10 nJ cm$^{-2}$ per pulse, $N_0 \sim 3\times10^{14}$ cm$^{-3}$) from the front (top surface) and back (glass side).

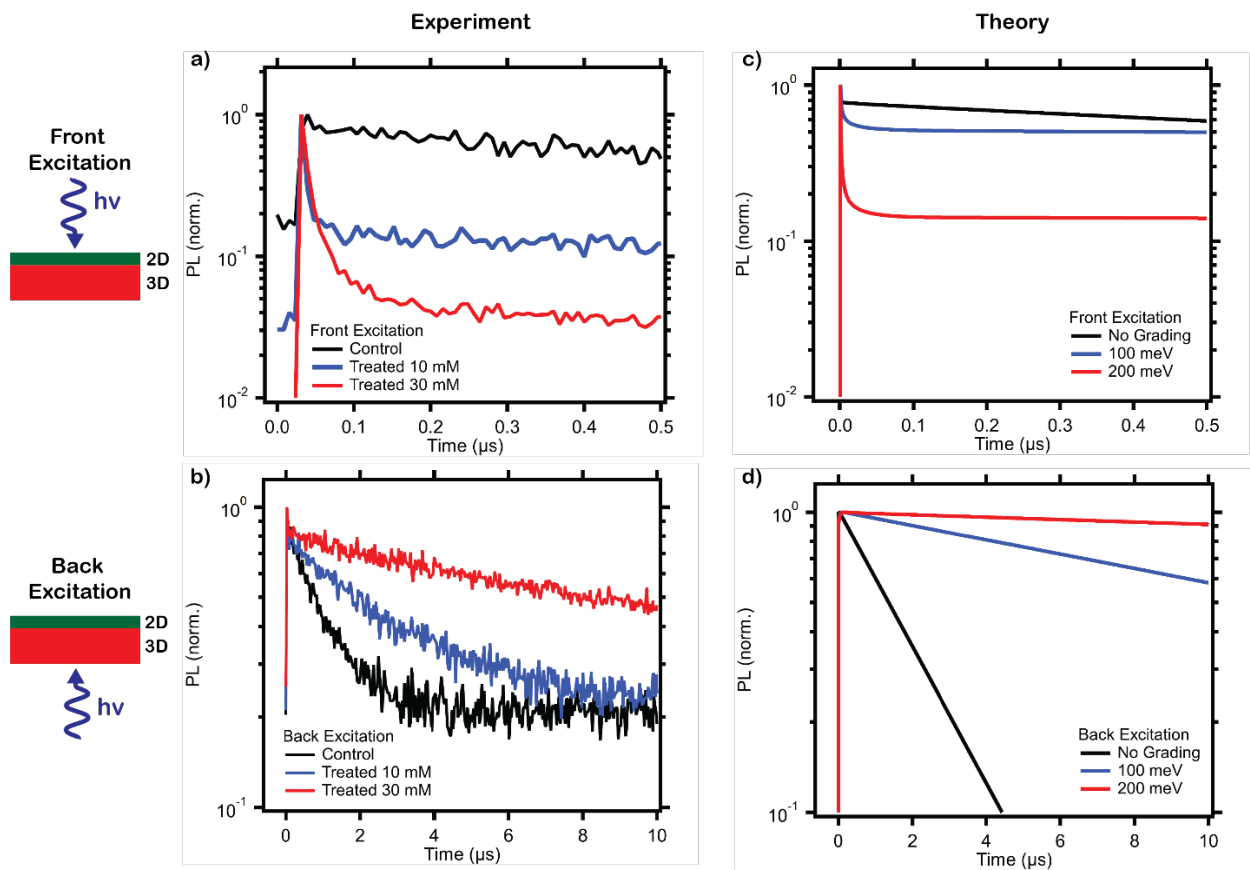

**Figure S23.** a) Experimental time-resolved photoluminescence (PL) decay traces when excited from the front (see schematics to left of panels) and b) back side of the 2D/3D heterostructure for a control (black trace), 10 mM HAbr treated (blue trace), and 30 mM HABr treated (red trace) samples. c) Simulated time-resolved PL decay traces when excited from the c) front and d) back side for a perovskite sample with no grading (black), 100 meV (blue), and 200 meV (red) gradings (i.e. difference between flat band and maximum at interface). The band diagrams for these simulations are shown in Figure S21a. For additional simulation details see section with header "Explanation of PL Behaviour of 2D/3D Perovskite Heterostructures", which contains all simulation parameters.

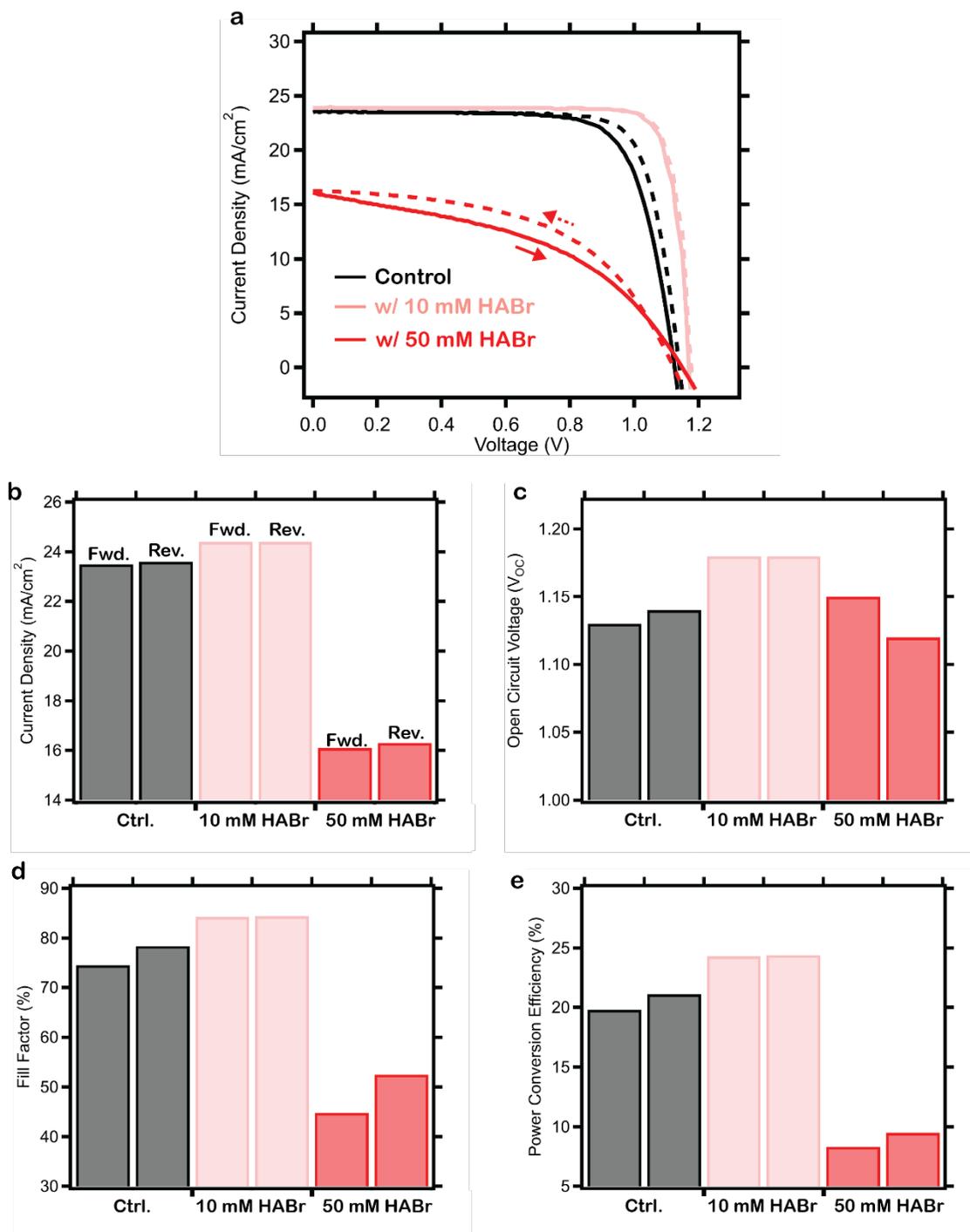

**Figure S24.** a) Current-voltage curves of control (black), 10 mM HABr treated (light red), and 50 mM HABr treated (red) solar cells. Summary of photovoltaic device parameters including b) current density, c) open circuit voltage, d) fill factor, and e) power conversion efficiency. The 10 mM device gives the best solar cell performance compared to the control and 50 mM HABr treated samples.

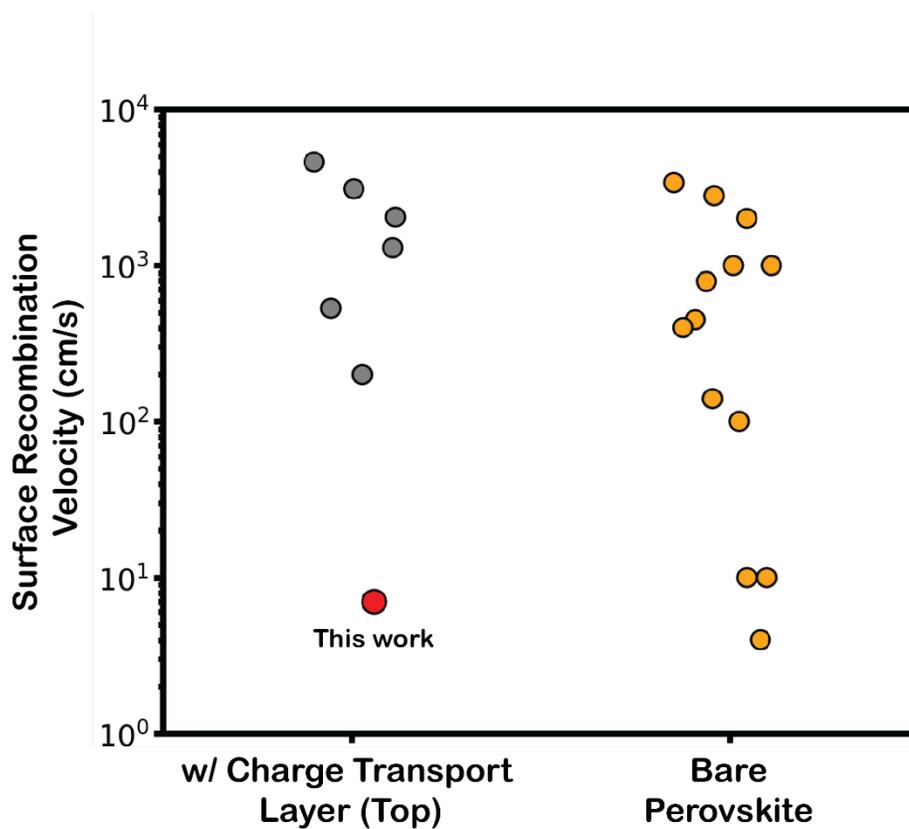

**Figure S25.** Literature survey of surface recombination velocity (SRV) of perovskites materials interfaced with a charge transport layer deposited on top of the perovskite compared to perovskite samples with no charge transport layer (i.e. "Bare Perovskite"). The SRV achieved in this work is comparable to the best SRV of passivated perovskites that have not been interfaced with any charge transport layer. See Supplementary Table S6 for references.

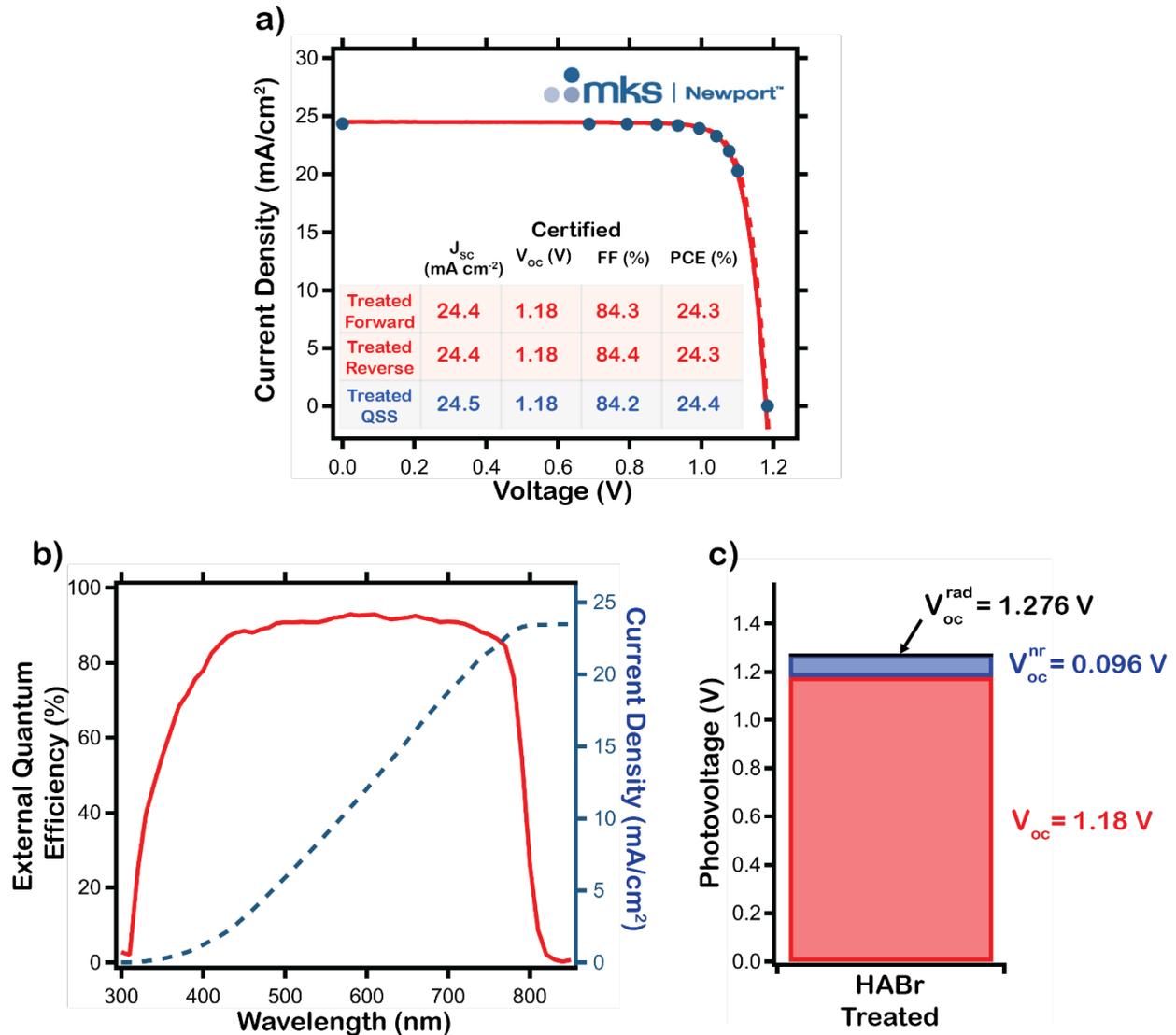

**Figure S26.** a) Newport certified current-voltage (JV) curves scanned in the forward and reverse directions as well as the quasi-steady state (QSS) JV curve, which is also shown in Figure 4d in the main article. b) External quantum efficiency (EQE) spectrum along with the cumulative sum of the current density ($J_{sc,sum}$ = 23.5 mA cm$^{-2}$). c) Radiative $V_{OC}$ ($V_{OC}^{rad}$) calculated using the EQE spectrum in b) as in input into equation S21. The non-radiative voltage loss ,$V_{OC}^{nr}$, is determined by subtracting $V_{OC}^{rad}$ from the device $V_{oc}$. The measured $V_{oc}$ is ~92% of the radiative limit.

| Reference | Publication Date | Lifetime (μs) |
|---|---|---|
| Kim, H.S. et al. Sci. Rep., 2, 591 (2012). | 2012.64 | 0.35 |
| Stranks, S.D. et al. Science, 342, 6156, 341-344, (2013). | 2013.80 | 0.273 |
| Xing, G. et al. Science, 342, 6156, 344-347, (2013). | 2013.80 | 0.0045 |
| Zhou, H. et al. Science, 345, 6196, 542-546, (2014) | 2014.58 | 0.382 |
| Zhou, H. et al. Science, 345, 6196, 542-546, (2014) | 2014.58 | 0.736 |
| Zhang, M. et al. Chem. Commun., 50, 11727-11730, (2014). | 2014.62 | 0.446 |
| Noel, N. et al. ACS Nano, 8, 10, 9815–9821, (2014). | 2014.66 | 0.3407 |
| Noel, N. et al. ACS Nano, 8, 10, 9815–9821, (2014). | 2014.66 | 2.0165 |
| Noel, N. et al. ACS Nano, 8, 10, 9815–9821, (2014). | 2014.66 | 2.2342 |
| You, J. et al. Appl. Phys. Lett. 105, 183902 (2014). | 2014.85 | 0.091 |
| Nie, W. et al. Science, 347, 6221, 522-525, (2015). | 2015.08 | 0.0096 |
| Wang, H.H. et al. J. Mater. Chem. A, 3, 9108-9115, (2015). | 2015.17 | 0.08 |
| Wang, H.H. et al. J. Mater. Chem. A, 3, 9108-9115, (2015). | 2015.17 | 0.081 |
| Wang, H.H. et al. J. Mater. Chem. A, 3, 9108-9115, (2015). | 2015.17 | 0.087 |
| Wang, H.H. et al. J. Mater. Chem. A, 3, 9108-9115, (2015). | 2015.17 | 0.096 |
| Wang, H.H. et al. J. Mater. Chem. A, 3, 9108-9115, (2015). | 2015.17 | 0.102 |
| Chen, Q. et al. Nat. Commun., 6, 7269, (2015). | 2015.45 | 0.241 |
| Chen, Q. et al. Nat. Commun., 6, 7269, (2015). | 2015.45 | 0.289 |
| Milot, R. L. et al. Adv. Funct. Mater., 25, 6218-6227, (2015). | 2015.71 | 0.0666 |
| Song, D. et al. J. Phys. Chem. C, 119, 40, 22812–22819, (2015). | 2015.72 | 0.044 |
| Song, D. et al. J. Phys. Chem. C. 119, 22812 (2015). | 2015.72 | 0.028 |
| Song, D. et al. J. Phys. Chem. C. 119, 22812 (2015). | 2015.72 | 0.045 |
| Li, Y. et al. J. Am. Chem. Soc. 137, 49, 15540-15547, (2015). | 2015.90 | 0.0632 |
| Li, Y. et al. J. Am. Chem. Soc. 137, 49, 15540-15547, (2015). | 2015.90 | 0.172 |
| deQuilettes, D.W. et al. ACS Energy Lett. 1, 438, (2016). | 2016.57 | 0.97 |
| deQuilettes, D.W. et al. ACS Energy Lett. 1, 438, (2016). | 2016.57 | 8.8 |
| Staub, F. et al. Phys. Rev. Appl. 6 (4), 1–13, (2016). | 2016.82 | 0.51 |
| Richter, J. et al. Nat. Commun., 7, 13941, (2016). | 2016.98 | 0.85 |
| Richter, J. M. et al. Nat. Commun. 7, 13941, (2016). | 2016.98 | 0.85 |
| Tan, H. et al. Science, 355 (6326), 722–726, (2017). | 2017.09 | 0.47 |
| Xue, Q. et al. Adv. Energy Mater., 7, 1602333, (2017). | 2017.36 | 0.0201 |
| Xue, Q. et al. Adv. Energy Mater., 7, 1602333, (2017). | 2017.36 | 0.0326 |
| Minh, et al. Chem. Mater., 29, 13, 5713–5719, (2017). | 2017.45 | 0.08017 |
| Yang, W. S. et al. Science 356 (6345), 1376–1379, (2017). | 2017.47 | 0.23 |
| Yang, W. S. et al. Science 356 (6345), 1376–1379, (2017). | 2017.47 | 1.1 |
| Zheng, X. et al. Nature Energy, 2, 17102, (2017). | 2017.49 | 0.9034 |
| Staub et al. J. Phys. Chem. Lett., 8, 20, 5084–5090, (2017). | 2017.76 | 0.87 |
| Turren-Cruz, S.-H. et al. Energy Environ. Sci. 11, 78, (2018). | 2017.93 | 0.055 |
| Turren-Cruz, S.-H. et al. Energy Environ. Sci. 11, 78, (2018). | 2017.93 | 1.33 |
| Turren-Cruz, S.-H. et al. Energy Environ. Sci. 11, 78, (2018). | 2017.93 | 1.37 |
| Turren-Cruz, S.-H. et al. Energy Environ. Sci. 11, 78, (2018). | 2017.93 | 1.93 |
| Kubicki, D. et al. J. Am. Chem. Soc., 140, 3345-3351, (2018). | 2018.12 | 0.081 |
| Kubicki, D. et al. J. Am. Chem. Soc., 140, 3345-3351, (2018). | 2018.12 | 3.1 |
| Abdi-Jalebi, M. et al. Nature, 555, 497, (2018) | 2018.22 | 0.45 |
| Braly, I.L.; D. W. deQuilettes, D.W. et al. Nat. Photonics, 12, 355, (2018). | 2018.33 | 0.35 |
| Braly, I.L.; D. W. deQuilettes, D.W. et al. Nat. Photonics, 12, 355, (2018). | 2018.33 | 5.5 |
| Du, T. et al. Adv. Funct. Mater., 28, 1801808, (2018). | 2018.48 | 0.37 |
| Du, T. et al. Adv. Funct. Mater., 28, 1801808, (2018). | 2018.48 | 0.735 |
| Abdi-Jalebi, M. et al. ACS Nano, 12 (7), 7301–7311, (2018). | 2018.49 | 0.16 |
| Abdi-Jalebi, M. et al. ACS Nano, 12 (7), 7301–7311, (2018). | 2018.49 | 0.2 |
| Jeon, N. J. et al. Nat. Energy, 3 (8), 682–689, (2018). | 2018.52 | 0.41 |
| Stolterfoht, M. et al. Nat. Energy, 3, 847, (2018). | 2018.58 | 0.026 |
| Stolterfoht, M. et al. Nat. Energy, 3, 847, (2018). | 2018.58 | 0.03 |
| Stolterfoht, M. et al. Nat. Energy, 3, 847, (2018). | 2018.58 | 0.18 |
| Stolterfoht, M. et al. Nat. Energy, 3, 847, (2018). | 2018.58 | 0.5 |
| Liu, Z. et al. ACS Energy Lett. 4, 110, (2019). | 2018.93 | 1.8 |
| Caprioglio, P. et al. Sustain. Energy Fuels 3, 550, (2019). | 2019.03 | 0.15 |
| Caprioglio, P. et al. Sustain. Energy Fuels 3, 550, (2019). | 2019.03 | 0.16 |
| Caprioglio, P. et al. Sustain. Energy Fuels 3, 550, (2019). | 2019.03 | 0.21 |
| Caprioglio, P. et al. Sustain. Energy Fuels 3, 550, (2019). | 2019.03 | 0.35 |
| Caprioglio, P. et al. Sustain. Energy Fuels 3, 550, (2019). | 2019.03 | 0.58 |
| Caprioglio, P. et al. Sustain. Energy Fuels 3, 550, (2019). | 2019.03 | 0.8 |
| Caprioglio, P. et al. Sustain. Energy Fuels 3, 550, (2019). | 2019.03 | 0.92 |
| Abdi-Jalebi, M. et al. Sci. Adv. 5, eaav2012, (2019). | 2019.13 | 0.16 |
| Abdi-Jalebi, M. et al. Sci. Adv. 5, eaav2012, (2019). | 2019.13 | 0.2 |
| Abdi-Jalebi, M. et al. Sci. Adv. 5, eaav2012, (2019). | 2019.13 | 0.4 |
| Alharbi, E. A. et al. Research. 2019, 8474698 (2019). | 2019.21 | 0.215 |
| Alharbi, E. A. et al. Research. 2019, 8474698 (2019). | 2019.21 | 0.627 |
| Wu, S. et al. Chem. Commun. 55, 4315, (2019). | 2019.21 | 0.771 |
| Wu, S. et al. Chem. Commun. 55, 4315, (2019). | 2019.21 | 1.22 |
| Wu, S. et al. Chem. Commun. 55, 4315, (2019). | 2019.21 | 1.35 |
| Wu, S. et al. Chem. Commun. 55, 4315, (2019). | 2019.21 | 1.85 |
| Wu, S. et al. Chem. Commun. 55, 4315, (2019). | 2019.21 | 2.66 |
| Yang, S. et al. J. Am. Chem. Soc. 141, 5781, (2019). | 2019.21 | 0.245 |
| Yang, S. et al. J. Am. Chem. Soc. 141, 5781, (2019). | 2019.21 | 0.267 |
| Yang, S. et al. J. Am. Chem. Soc. 141, 5781, (2019). | 2019.21 | 0.491 |
| Yang, S. et al. J. Am. Chem. Soc. 141, 5781, (2019). | 2019.21 | 0.556 |
| Yang, S. et al. J. Am. Chem. Soc. 141, 5781, (2019). | 2019.21 | 0.737 |
| Yang, S. et al. J. Am. Chem. Soc. 141, 5781, (2019). | 2019.21 | 0.781 |
| Yang, S. et al. J. Am. Chem. Soc. 141, 5781, (2019). | 2019.21 | 0.93 |
| Yang, S. et al. J. Am. Chem. Soc. 141, 5781, (2019). | 2019.21 | 1.28 |
| Jiang, Q. et al. Nat. Photonics, 13 (7), 460–466, (2019). | 2019.25 | 0.364 |
| Jiang, Q. et al. Nat. Photonics, 13 (7), 460–466, (2019). | 2019.25 | 2.825 |
| Tan, F. et al. Adv. Mater. 31, 14, 1807435, (2019) | 2019.26 | 1.0295 |
| Gong, W. et al. Nat. Commun., 10, 1591, (2019). | 2019.27 | 0.981 |
| Tong, J. et al. Science, 364 (6439), 475–479, (2019). | 2019.30 | 0.14 |
| Tong, J. et al. Science, 364 (6439), 475–479, (2019). | 2019.30 | 1.2 |
| Wang, R. et al. Joule, 3, 1464-1477, (2019). | 2019.47 | 0.0519 |
| Wang, R. et al. Joule, 3, 1464-1477, (2019). | 2019.47 | 0.1143 |
| Kim, M. et al. Joule, 3 (9), 2179–2192, (2019). | 2019.47 | 0.33 |
| Kim, M. et al. Joule, 3 (9), 2179–2192, (2019). | 2019.47 | 1.4 |
| Wang, Q. et al. Adv. Energy Mater. 9, 1900990, (2019). | 2019.48 | 0.026 |
| Wang, Q. et al. Adv. Energy Mater. 9, 1900990, (2019). | 2019.48 | 0.103 |
| Wang, Q. et al. Adv. Energy Mater. 9, 1900990, (2019). | 2019.48 | 0.24 |
| Wang, Q. et al. Adv. Energy Mater. 9, 1900990, (2019). | 2019.48 | 0.343 |
| Yang, S. et al. Science 365, 6452, 473-478, (2019) | 2019.59 | 0.12 |
| Yang, S. et al. Science 365, 6452, 473-478, (2019) | 2019.59 | 0.49 |
| Bowman, A. R. et al. ACS Energy Lett., 4, 2301-2307, (2019) | 2019.64 | 0.37 |
| Bowman, A. R. et al. ACS Energy Lett., 4, 2301-2307, (2019) | 2019.64 | 1.17 |
| Bowman, A. R. et al. ACS Energy Lett., 4, 2301-2307, (2019) | 2019.64 | 1.19 |
| Bowman, A. R. et al. ACS Energy Lett., 4, 2301-2307, (2019) | 2019.64 | 1.52 |
| Al-Ashouri, A. et al. Energy Environ. Sci. 12, 3356, (2019). | 2019.75 | 0.11 |
| Al-Ashouri, A. et al. Energy Environ. Sci. 12, 3356, (2019). | 2019.75 | 0.15 |
| Al-Ashouri, A. et al. Energy Environ. Sci. 12, 3356, (2019). | 2019.75 | 0.6 |
| Al-Ashouri, A. et al. Energy Environ. Sci. 12, 3356, (2019). | 2019.75 | 0.75 |
| Min, H. et al. Science, 366 (6466), 749–753, (2019). | 2019.85 | 0.72 |
| Min, H. et al. Science, 366 (6466), 749–753, (2019). | 2019.85 | 1.6 |
| Wang, R. et al. Science, 366, 6472, (2019) | 2019.97 | 0.64 |
| Wang, J. et al. Nat. Commun. 11, 177, (2020) | 2020.03 | 0.0075 |
| Wang, J. et al. Nat. Commun. 11, 177, (2020) | 2020.03 | 0.0121 |
| Zheng, X. et al. Nat. Energy, 5 (2), 131–140, (2020). | 2020.05 | 0.11 |
| Zheng, X. et al. Nat. Energy, 5 (2), 131–140, (2020). | 2020.05 | 0.79 |
| Zheng, X. et al. Nat. Energy, 5 (2), 131–140, (2020). | 2020.05 | 1.1 |
| Back, H. et al. Energy Environ. Sci. 13, 840-847, (2020). | 2020.13 | 0.01 |
| Back, H. et al. Energy Environ. Sci. 13, 840-847, (2020). | 2020.13 | 0.099 |
| Back, H. et al. Energy Environ. Sci. 13, 840-847, (2020). | 2020.13 | 0.112 |
| Hou, Y. et al. Science, 367 (6482), 1135 LP – 1140, (2020). | 2020.18 | 0.57 |
| Chen, B. et al.  Nat. Commun., 11, 1257, (2020). | 2020.19 | 1 |
| Chen, B. et al.  Nat. Commun., 11, 1257, (2020). | 2020.19 | 1.732 |
| Ansari, F. et al. J. Am. Chem. Soc., 142, 26, 11428-11433, (2020). | 2020.36 | 2.8 |
| Tian, J. et al. Adv. Funct. Mater., *30,* 2001764, (2020). | 2020.42 | 0.00068 |
| Tian, J. et al. Adv. Funct. Mater., *30,* 2001764, (2020). | 2020.42 | 0.00124 |
| Wu, S. et al. Joule, 4, 6, 1248-1262, (2020) | 2020.46 | 0.2577 |
| Yao, Q. et al Adv. Mater. 32, 2000571, (2020). | 2020.48 | 8.37 |
| Yao, Q. et al Adv. Mater. 32, 2000571, (2020). | 2020.48 | 11.98 |
| Du, T. et al. J. Mater. Chem. C, 8, 12648-12655, (2020). | 2020.64 | 0.12 |
| Du, T. et al. J. Mater. Chem. C, 8, 12648-12655, (2020). | 2020.64 | 0.13 |
| Xiao, K. et al. Nat. Energy, 5, 870-880, (2020). | 2020.76 | 0.134 |
| Xiao, K. et al. Nat. Energy, 5, 870-880, (2020). | 2020.76 | 0.631 |
| Wolff, C. M. et al. ACS Nano, 14 (2), 1445–1456, (2020). | 2020.86 | 0.8 |
| Al-Ashouri, A. et al. Science, 370, 6522, 1300-1309, (2020). | 2020.95 | 5 |
| Gutierrez-Partida, E. et al. ACS Energy Lett. 6, 1045-1054 (2021). | 2021.13 | 1 |
| Gutierrez-Partida, E. et al. ACS Energy Lett. 6, 1045-1054 (2021). | 2021.13 | 18.2 |
| Jariwala, S. et al. Chem. Mater., 33, 13, 5035-5044, (2021) | 2021.46 | 0.6 |
| Jariwala, S. et al. Chem. Mater., 33, 13, 5035-5044, (2021) | 2021.46 | 4.3 |
| Chen, B. et al. Adv. Mater. 33, 2103394, (2021). | 2021.65 | 0.3564 |
| Ding, Y. et al. Nat. Nanotechnol., 17, 598-605, (2022). | 2022.31 | 0.093 |

**Table S2.** Literature survey of perovskite charge carrier lifetimes determined from standard time-resolved photoluminescence measurements.

| Reference | Publication Date | Lifetime (µs) |
|---|---|---|
| Hwang, C.J. J. Appl. Phys., 42, 4408, (1971). | 1971.75 | 0.02 |
| Nelson, R.J. et al. J. Appl. Phys., 49, 6103, (1978). | 1978.92 | 1.3 |
| Olson, J.M. et al. Appl. Phys. Lett., 55, 1208, (1989). | 1989.53 | 14.2 |
| Olson, J.M. et al. Appl. Phys. Lett., 55, 1208, (1989). | 1989.53 | 8.5 |
| Olson, J.M. et al. Appl. Phys. Lett., 55, 1208, (1989). | 1989.53 | 0.29 |
| Olson, J.M. et al. Appl. Phys. Lett., 55, 1208, (1989). | 1989.53 | 0.69 |
| Olson, J.M. et al. Appl. Phys. Lett., 55, 1208, (1989). | 1989.53 | 0.055 |
| Ahrenkiel, R.K. et al. Appl. Phys. Lett., 55, 1088, (1989). | 1989.51 | 0.153 |
| Ahrenkiel, R.K. et al. Appl. Phys. Lett., 55, 1088, (1989). | 1989.51 | 0.19 |
| Molenkamp, L.W. et al. Appl. Phys. Lett., 54, 1992, (1989). | 1989.18 | 4.9 |
| Yablonovitch, E. et al. Appl. Phys. Lett., 50, 1197, (1987). | 1987.25 | 2 |
| Yablonovitch, E. et al. Appl. Phys. Lett., 50, 1197, (1987). | 1987.25 | 1 |
| Yablonovitch, E. et al. Appl. Phys. Lett., 50, 1197, (1987). | 1987.25 | 0.5 |
| Smith, L.M. et al. J. Vac. Sci. Technol. B., 8, 787, (1990). | 1990.3 | 0.804 |
| Hooft, G.W.'t et al. Jpn. J. Appl. Phys., 24, L761, (1985). | 1985.67 | 1.63 |
| Hooft, G.W.'t et al. Jpn. J. Appl. Phys., 24, L761, (1985). | 1985.67 | 0.19 |
| Gilliland, G.D. et al. Appl. Phys. Lett., 59, 216, (1991). | 1991.27 | 2.5 |
| Hummel, S.G. et al. Appl. Phys. Lett., 57, 695, (1990). | 1990.41 | 0.4 |
| Dawson, P. et al. Appl. Phys. Lett., 45, 1227, (1984). | 1984.72 | 0.0033 |
| Dawson, P. et al. Appl. Phys. Lett., 45, 1227, (1984). | 1984.72 | 0.09 |
| Dawson, P. et al. Appl. Phys. Lett., 45, 1227, (1984). | 1984.72 | 0.0049 |
| Dawson, P. et al. Appl. Phys. Lett., 45, 1227, (1984). | 1984.72 | 0.056 |
| Dawson, P. et al. Appl. Phys. Lett., 45, 1227, (1984). | 1984.72 | 0.011 |
| Dawson, P. et al. Appl. Phys. Lett., 45, 1227, (1984). | 1984.72 | 0.125 |
| Dawson, P. et al. Appl. Phys. Lett., 45, 1227, (1984). | 1984.72 | 0.024 |
| Ahrenkiel, R. K. Semiconductors and Semimetals 39, 39-150, (1993). | 1993 | 1.08 |
| Haughn, C.R. et al. Appl. Phys. Lett., 102, 182108, (2013). | 2013.36 | 1.6 |
| Johnson, S.R. et al. J. Vac. Sci. Technol. B., 25, 1077, (2007). | 2007.41 | 0.0599 |
| Oshima, R. et al. IEEE J. Photovolt., 9, 1, (2019). | 2018.84 | 0.136 |

**Table S3.** Literature survey of gallium arsenide (GaAs) charge carrier lifetimes determined from standard time-resolved photoluminescence measurements.

| Reference | Publication Date | Lifetime (µs) |
|---|---|---|
| Yamaguchi, M. et al. J. Appl. Phys., 52, 6429, (1981). | 1981.45 | 3.003 |
| Yamaguchi, M. et al. J. Appl. Phys., 52, 6429, (1981). | 1981.45 | 0.356 |
| Ahrenkiel, R.K. et al. Solar Cells, 24, 3-4, 339-353, (1988). | 1988.59 | 0.21 |
| Ahrenkiel, R.K. et al. Solar Cells, 24, 3-4, 339-353, (1988). | 1988.59 | 0.303 |
| Bhimnathwala, H.G. et al. Proc.21st lEEE PVSC, 394, (1990). | 1990.39 | 0.035 |
| Bhimnathwala, H.G. et al. Proc.21st lEEE PVSC, 394, (1990). | 1990.39 | 0.006 |
| Bhimnathwala, H.G. et al. Proc.21st lEEE PVSC, 394, (1990). | 1990.39 | 0.005 |
| Bothra, S. et al. Proc. 21st IEEE PVSC, 404, (1990). | 1990.39 | 0.039 |
| Bothra, S. et al. Proc. 21st IEEE PVSC, 404, (1990). | 1990.39 | 0.029 |
| Bothra, S. et al. Proc. 21st IEEE PVSC, 404, (1990). | 1990.39 | 0.020 |
| Landis, G.A. et al. Proc. 3rd Int. Conf. InP and Rel. Compounds, 636, (1991). | 1991.42 | 0.089 |
| Landis, G.A. et al. Proc. 3rd Int. Conf. InP and Rel. Compounds, 636, (1991). | 1991.42 | 0.061 |
| Landis, G.A. et al. Proc. 3rd Int. Conf. InP and Rel. Compounds, 636, (1991). | 1991.42 | 0.043 |
| Landis, G.A. et al. Proc. 3rd Int. Conf. InP and Rel. Compounds, 636, (1991). | 1991.42 | 0.022 |
| Landis, G.A. et al. Proc. 3rd Int. Conf. InP and Rel. Compounds, 636, (1991). | 1991.42 | 0.013 |
| Landis, G.A. et al. Proc. 3rd Int. Conf. InP and Rel. Compounds, 636, (1991). | 1991.42 | 0.010 |
| Landis, G.A. et al. Proc. 3rd Int. Conf. InP and Rel. Compounds, 636, (1991). | 1991.42 | 0.008 |
| Landis, G.A. et al. Proc. 3rd Int. Conf. InP and Rel. Compounds, 636, (1991). | 1991.42 | 0.494 |
| Landis, G.A. et al. Proc. 3rd Int. Conf. InP and Rel. Compounds, 636, (1991). | 1991.42 | 0.332 |
| Landis, G.A. et al. Proc. 3rd Int. Conf. InP and Rel. Compounds, 636, (1991). | 1991.42 | 0.300 |
| Landis, G.A. et al. Proc. 3rd Int. Conf. InP and Rel. Compounds, 636, (1991). | 1991.42 | 0.250 |
| Landis, G.A. et al. Proc. 3rd Int. Conf. InP and Rel. Compounds, 636, (1991). | 1991.42 | 0.473 |
| Landis, G.A. et al. Proc. 3rd Int. Conf. InP and Rel. Compounds, 636, (1991). | 1991.42 | 0.421 |
| Landis, G.A. et al. Proc. 3rd Int. Conf. InP and Rel. Compounds, 636, (1991). | 1991.42 | 0.248 |
| Landis, G.A. et al. Proc. 3rd Int. Conf. InP and Rel. Compounds, 636, (1991). | 1991.42 | 0.161 |
| Jenkins, P. et al. 22nd IEEE PVSC, 4243892, (1991). | 1991.75 | 0.7 |
| Jenkins, P. et al. 22nd IEEE PVSC, 4243892, (1991). | 1991.75 | 0.23 |
| Rosenwaks, Y. et al. Phys. Rev., 44, 13098, (1991). | 1991.96 | 0.023 |
| Rosenwaks, Y. et al. Phys. Rev., 44, 13098, (1991). | 1991.96 | 0.004 |
| Rosenwaks, Y. et al. Phys. Rev., 44, 13098, (1991). | 1991.96 | 0.332 |
| Keyes, B.M. et al. J. Appl. Phys., 75, 4249, (1994). | 1993.97 | 3.17 |
| Liu, A. et al. J. Appl. Phys., 86, 430, (1990). | 1999.21 | 0.073 |
| Liu, A. et al. J. Appl. Phys., 86, 430, (1990). | 1999.21 | 0.088 |
| Liu, A. et al. J. Appl. Phys., 86, 430, (1990). | 1999.21 | 0.133 |
| Liu, A. et al. J. Appl. Phys., 86, 430, (1990). | 1999.21 | 0.142 |
| Liu, A. et al. J. Appl. Phys., 86, 430, (1990). | 1999.21 | 0.084 |
| Liu, A. et al. J. Appl. Phys., 86, 430, (1990). | 1999.21 | 0.088 |
| Liu, A. et al. J. Appl. Phys., 86, 430, (1990). | 1999.21 | 0.113 |
| Liu, A. et al. J. Appl. Phys., 86, 430, (1990). | 1999.21 | 0.114 |
| Johnston, S. et al. IEEE 42nd PVSC., 15664411, (2015). | 2015.96 | 0.012 |

**Table S4.** Literature survey of indium phosphide (InP) charge carrier lifetimes determined from standard time-resolved photoluminescence measurements.

| Reference | Publication Date | Lifetime (µs) |
|---|---|---|
| Gessert, T.A. et al. 37th IEEE PVSC., (2011). | 2011.48 | 0.0023 |
| Barnard, E.S. et al. Sci. Rep., 3, 2098, (2013). | 2013.49 | 0.0007 |
| Barnard, E.S. et al. Sci. Rep., 3, 2098, (2013). | 2013.49 | 0.0018 |
| Barnard, E.S. et al. Sci. Rep., 3, 2098, (2013). | 2013.49 | 0.0054 |
| Barnard, E.S. et al. Sci. Rep., 3, 2098, (2013). | 2013.49 | 0.0092 |
| Barnard, E.S. et al. Sci. Rep., 3, 2098, (2013). | 2013.49 | 0.0177 |
| Barnard, E.S. et al. Sci. Rep., 3, 2098, (2013). | 2013.49 | 0.073 |
| Ma, J. et al. Phys. Rev. Lett., 111, 067402, (2013) | 2013.6 | 0.0034 |
| Ma, J. et al. Phys. Rev. Lett., 111, 067402, (2013) | 2013.6 | 0.0196 |
| Ma, J. et al. Phys. Rev. Lett., 111, 067402, (2013) | 2013.6 | 0.009 |
| Ma, J. et al. Phys. Rev. Lett., 111, 067402, (2013) | 2013.6 | 0.0025 |
| Ma, J. et al. Phys. Rev. Lett., 111, 067402, (2013) | 2013.6 | 0.0015 |
| Swartz, C.H. et al. Appl. Phys. Lett., 105, 222107, (2014). | 2014.92 | 0.24 |
| Zhao, X.-H. et al. Appl. Phys. Lett., 105, 252101, (2014). | 2014.98 | 0.179 |
| Zhao, X.-H. et al. Appl. Phys. Lett., 105, 252101, (2014). | 2014.98 | 0.086 |
| Zhao, X.-H. et al. Appl. Phys. Lett., 105, 252101, (2014). | 2014.98 | 0.042 |
| Zhao, X.-H. et al. Appl. Phys. Lett., 105, 252101, (2014). | 2014.98 | 0.031 |
| Reese, M.O. et al. J. Appl. Phys., 118, 155305, (2015). | 2015.81 | 0.14 |
| Reese, M.O. et al. J. Appl. Phys., 118, 155305, (2015). | 2015.81 | 0.014 |
| Burst, J.M. et al. APL Materials, 4, 116102, (2016). | 2016.88 | 0.0002 |
| Burst, J.M. et al. APL Materials, 4, 116102, (2016). | 2016.88 | 0.0002 |
| Burst, J.M. et al. APL Materials, 4, 116102, (2016). | 2016.88 | 0.0001 |
| Burst, J.M. et al. APL Materials, 4, 116102, (2016). | 2016.88 | 0.0002 |
| Burst, J.M. et al. APL Materials, 4, 116102, (2016). | 2016.88 | 0.0051 |
| Burst, J.M. et al. APL Materials, 4, 116102, (2016). | 2016.88 | 0.0069 |
| Burst, J.M. et al. APL Materials, 4, 116102, (2016). | 2016.88 | 0.0049 |
| Burst, J.M. et al. APL Materials, 4, 116102, (2016). | 2016.88 | 0.0782 |
| Burst, J.M. et al. APL Materials, 4, 116102, (2016). | 2016.88 | 0.0907 |
| Burst, J.M. et al. APL Materials, 4, 116102, (2016). | 2016.88 | 0.0388 |
| Burst, J.M. et al. APL Materials, 4, 116102, (2016). | 2016.88 | 0.0196 |
| Burst, J.M. et al. APL Materials, 4, 116102, (2016). | 2016.88 | 0.0766 |
| Burst, J.M. et al. APL Materials, 4, 116102, (2016). | 2016.88 | 0.0449 |
| Burst, J.M. et al. APL Materials, 4, 116102, (2016). | 2016.88 | 0.0205 |
| Burst, J.M. et al. APL Materials, 4, 116102, (2016). | 2016.88 | 0.0386 |
| Burst, J.M. et al. APL Materials, 4, 116102, (2016). | 2016.88 | 0.0231 |
| Ablekin, T. et al. Solar RRL, 5, 8, 2100173, (2021). | 2021.37 | 0.256 |

**Table S5.** Literature survey of cadmium telluride (CdTe) charge carrier lifetimes determined from standard time-resolved photoluminescence measurements.

| Reference | Surface Recombination Velocity (cm/s) | Charge Transport Layer Deposited on Top |
|---|---|---|
| Wang, J. et al. ACS Energy Lett., 4, 1, 222-227, (2019). | 3100 | Spiro-OMeTAD |
| Wang, J. et al. ACS Energy Lett., 4, 1, 222-227, (2019). | 532 | TPBi |
| Wang, J. et al. ACS Energy Lett., 4, 1, 222-227, (2019). | 1300 | ITIC |
| Wang, J. et al. ACS Energy Lett., 4, 1, 222-227, (2019). | 2040 | ICBA |
| Wang, J. et al. ACS Energy Lett., 4, 1, 222-227, (2019). | 4600 | PCBM |
| Krogmeier, B. et al. Sustain. Energy Fuels, 2, 1027-1034, (2018) | 200 | PCBM |
| Wang, J. et al. ACS Energy Lett., 4, 1, 222-227, (2019). | 1000 | None |
| Wang, J. et al. ACS Energy Lett., 4, 1, 222-227, (2019). | 10 | None |
| Yang, Y. et al. Nat. Commun., 6, 7961, (2015). | 3400 | None |
| Yang, Y. et al. Nat. Energy, 2, 16207, (2017). | 2800 | None |
| Yang, Y. et al. Nat. Energy, 2, 16207, (2017). | 450 | None |
| Vidon, G. et al. Phys. Rev. Appl., 16, 044058, (2021). | 790 | None |
| Bercegol, A. et al. J. Phys. Chem. C, 122, 43, 24570-24577, (2018). | 100 | None |
| Bercegol, A. et al. J. Phys. Chem. C, 122, 43, 24570-24577, (2018). | 140 | None |
| Jariwala, S. et al. Chem. Mater., 33, 13, 5035-5044, (2021). | 1000 | None |
| Jariwala, S. et al. Chem. Mater., 33, 13, 5035-5044, (2021). | 10 | None |
| Jariwala, S. et al. Chem. Mater., 33, 13, 5035-5044, (2021). | 1 | None |
| Fang, H.-H. et al. Sci. Adv., 2, 7, (2016). | 4 | None |
| Wu, B. et al. Adv. Funct. Mater., 27, 7, 1604818, (2017). | 2000 | None |
| Gunes, U. et al. Adv. Funct. Mater., 31, 42, 2103130, (2021). | 400 | None |

**Table S6.** Literature survey of reported perovskite surface recombination velocities for charge transport layers (CTL's) deposited on top of the perovskite as well as bare (i.e. None) perovskite thin films or single crystals.